\documentclass[10pt]{iopart}
\usepackage{graphicx}
\usepackage{amsfonts}
\usepackage{amsbsy}
\usepackage{bbm}
\usepackage{dsfont}
\usepackage{color}
\usepackage{setstack}
\usepackage{epstopdf, epsfig}

\eqnobysec %equation numbering by section

\newcommand{\bra}{\langle}
\newcommand{\ket}{\rangle}
\newcommand{\bnull}{{\bf 0}}

\newcommand{\bb}{\boldsymbol{\beta}}

\newcommand{\tbb}{\tilde{\boldsymbol{\beta}}}
\newcommand{\tb}{\tilde{\beta}}

\newcommand{\uu}{\mathbf{u}}
\newcommand{\vv}{\mathbf{v}}

\newcommand{\y}{\mathbf{y}}

\newcommand{\z}{\mathbf{z}}

\newcommand{\A}{\boldsymbol{A}}

\newcommand{\C}{\boldsymbol{C}}

\newcommand{\D}{\boldsymbol{D}}

\newcommand{\bXi}{\boldsymbol{\Xi}}
\newcommand{\btXi}{\boldsymbol{\Xi}}

\newcommand{\Chat}{\hat{C}}

\newcommand{\I}{\mathbb{I}_p}

\newcommand{\PP}{\boldsymbol{P}}
\newcommand{\QQ}{\boldsymbol{Q}}

\newcommand{\param}{\boldsymbol{\vartheta}}
\newcommand{\lam}{\lambda}
\newcommand{\data}{\mathcal{D}}
\newcommand{\half}{\frac{1}{2}}

\newcommand{\zz}{\zeta}
\newcommand{\bxi}{\boldsymbol{\xi}}
\newcommand{\bomega}{\boldsymbol{\omega}}
\newcommand{\bOmega}{\boldsymbol{\Omega}}

\newcommand{\tS}{\tilde{S}}

\newcommand{\vsp}{\vspace*{3mm}}
\newcommand{\hsp}{\hspace*{3mm}}

\newcommand{\black}{\color{black}}

\begin{document}

\title[Analysis of overfitting in the penalized Cox model]{Analysis of  overfitting in the regularized Cox model} 

\author{Mansoor Sheikh$^{\dag\ddag}$ and Anthony CC Coolen$^{\dag\ddag\S}$}
\address{$\dag$Institute for Mathematical and Molecular Biomedicine, King's College London, Hodgkin Building, London SE1 1UL, UK\\
$\ddag$ Department of Mathematics,  King's College London, The Strand, London WC2R 2LS, UK \\
$\S$ Saddle Point Science Ltd, London, UK}

%check pacs numbers!!!
%\pacs{75.10.Nr,  02.50.Tt}
% 05.50.+q	Lattice theory and statistics (Ising, Potts, etc.) 
%05.70.Fh	Phase transitions: general studies
%02.50.-r	Probability theory, stochastic processes, and statistics
%02.50.Tt	 Inference methods
%75.10.Nr	Spin-glass and other random models (for spin glasses and other random magnets, see 75.50.Lk)

\ead{mansoor.sheikh@kcl.ac.uk, ton.coolen@kcl.ac.uk}

\begin{abstract}
The Cox proportional hazards model is ubiquitous in the analysis of time-to-event data. However, when the data dimension $p$ is comparable to the sample size $N$, maximum likelihood estimates for its regression parameters are known to be biased or break down entirely due to overfitting. This prompted the introduction of  the so-called regularized Cox model.  In this paper we use the replica method from statistical physics to investigate the relationship between the true and inferred regression parameters in regularized multivariate Cox regression with $L_2$ regularization, in the regime where both $p$ and $N$ are large but with $\zz=p/N \sim \mathcal{O}(1)$. We thereby  generalize a recent study  from maximum likelihood to maximum a posteriori inference. We also establish a relationship between the optimal regularization parameter and  $\zeta$, allowing for straightforward overfitting corrections in time-to-event analysis.
\end{abstract}

\noindent{\it Keywords\/}:  Cox proportional hazards, survival analysis, overfitting, MAP estimate, ridge regularization, replica method
%\maketitle\nopagebreak

%\tableofcontents\clearpage

\section{Introduction\label{section:Intro}}
%maximum likelihood, correction terms and regularization

Inference of parameters for generalized linear models using the maximum likelihood (ML) protocol becomes increasingly biased due to overfitting as the ratio $\zeta = p/N$ increases, where $p$ is the number of covariates and $N$ the number of training data. Overfitting occurs when model parameters seek to explain not only the `signal' but also the `noise' in training data, and is  characterized by a difference in outcome prediction accuracy between training and validation samples. See e.g. \cite{coolen2017replica,anderson1979logistic,sur2018modern,giles2009bias} for examples from logistic regression, gamma distributions and Cox models \cite{coolen2017replica}. 
Hence,  standard statistical significance tests for regression coefficients, being usually based on asymptotic results derived for fixed $p$, become increasingly inaccurate \cite{fan2017nonuniformity}. 
Unfortunately, in post-genome medicine, having large ratios $\zeta$ is the rule rather than the exception. This prompted epidemiologists to formulate heuristic rules for avoiding overfitting, such as limits on the number of events per variable \cite{concato1995importance,peduzzi1995importance,vittinghoff2007relaxing,courvoisier2011performance,greenland2016sparse}. The Cox proportional hazards model \cite{cox1972regression}, commonly used in epidemiological studies and clinical trials, predicts the continuous time-to-event random variable by combining an unspecified baseline hazard rate with a function of patient covariates. The canonical form for the covariate-dependent hazard rate of this model is $\lambda(t) = \lambda_0(t) e^{\bb.\z}$. It was originally developed for use with life-tables where $N$ is large (population-wide data) and the number of covariates $p$ is small. Maximizing a likelihood function is a valid inference method in this regime.

Approximate recipes for correcting ML estimates were developed in e.g. \cite{firth1993bias,cordeiro1991bias}. Alternative methods of addressing the overfitting problem include feature selection and regularization. 
In feature selection one seeks to identify a subset of covariates that are informative of outcomes \cite{fan2010selective,battey2018large,george1997approaches}. Its advantages include reduction in the required computational resources, and increased interpretability.  In regularization one adds a penalty term to the objective function of ML inference (which can alternatively be derived from a prior in Bayesian inference) to  suppress the number or magnitude of the model parameters \cite{friedman2009glmnet,tibshirani1996regression}. Application of regularization to survival analysis with high-dimensional covariates is   studied widely, see e.g.  \cite{witten2010survival,huang2002penalized} and references therein.

\begin{figure}[t]
\vspace*{-3mm}
\begin{center}\hspace*{6mm}
\includegraphics[width=10.5cm]{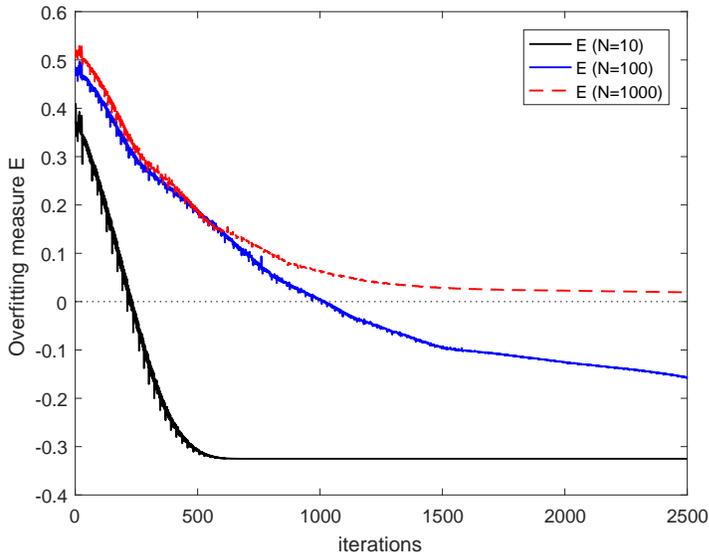} 
\end{center}
\vspace*{-3mm}
\caption{\small{Synthetic data with dimension $p\!=\!25$ and $N=\{10,100,1000 \}$ are generated using the logistic regression model. The ML estimate of model parameters is found numerically using the Nelder-Mead algorithm.  The overfitting measure $E$ is plotted after each iteration. The starting value model parameters $\bb$ in the minimization search   is the zero vector, giving a positive value of E (implying an underfitted model).}   } 
\label{fig:ML_intuition}
\end{figure}

\begin{figure}[t]
\vspace*{-3mm}
\begin{center}\hspace*{6mm}
\includegraphics[width=10.5cm]{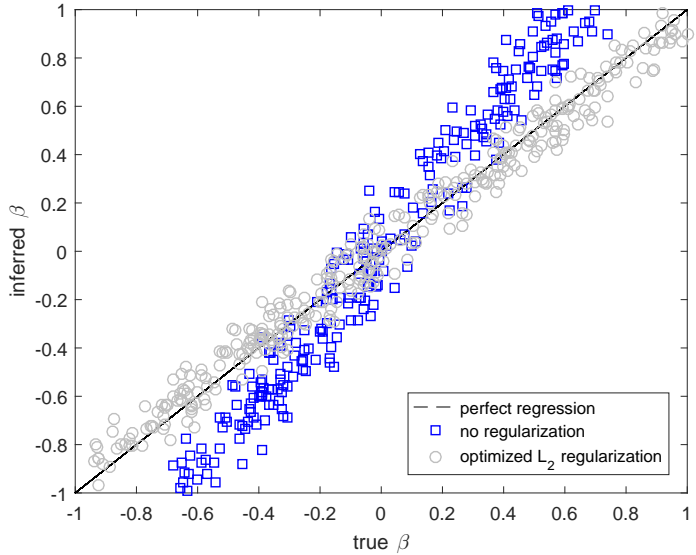} 
\end{center}
\vspace*{-3mm}
\caption{\small{Comparison of true and inferred regression coefficients for the Cox proportional hazards model. A systematic bias is found for non-zero values of $\zz$, which can be corrected with regularization. Synthetic survival data were generated \cite{therneau2017package} using Gaussian covariates ($p=500, ~N=833, ~\zz=0.6$). The regression coefficients are inferred \cite{friedman2009glmnet} using ML (no regularization) or Maximum A Posteriori Probability regression (MAP, with regularization).  Data points on the diagonal imply perfect inference. }
} 
\label{fig:slope}
\end{figure}

A recent study   \cite{coolen2017replica} provided a new approach to overfitting in survival analysis. It showed how the replica method from statistical physics can be used to  model  ML inference analytically as the zero noise limit of a suitably defined stochastic minimization, starting from an information-theoretic measure of overfitting.  The theory predicted the quantitative relation between ML-inferred and true parameters in the Cox model \cite{coolen2017replica}, and a phase transition at $\zeta=1$.

Let us denote the set of model parameters as $\param$, and the data as $\data$. The observation  that ML inference is equivalent to 
minimization of the Kullback-Leibler divergence between the empirical data distribution $\hat{P}_{\data} $ and the parametrized distribution $P_{\param} $ assumed as a model of the data, suggests  \cite{coolen2017replica}  using $E(\param, \data) \equiv D(\hat{P}_{\data}  \|  P_{\param} )  -  D(\hat{P}_{\data} \|  P_{\param^\star} ) $ as a measure of overfitting\footnote{A similar idea for comparing estimators of probability distributions was used in \cite{el2008spectrum}, using the L\'{e}vy distance rather than the KL divergence. Other measures of overfitting can be found in e.g. \cite{huang2002penalized,fan2013tuning}.}, in which $ \param^\star$ are the true (but a priori unknown) parameter values. Perfect regression implies $E=0$, underfitting implies $E>0$, and overfitting implies $E<0$.
To gain more intuition for this measure, we generate synthetic data  from a simple  logistic regression model, find the ML estimators of its parameters, and calculate $E$. Here the parameters are $\{ \bb_{\mu} \}_{\mu=0}^p$, the data are $\data=\{(t_1,\z_1),\ldots,(t_N,\z_N)\}$, with $\z_i\in \mathbb{R}^{p+1}$ and $t_i \in \{0,1\}$, and we use the short-hand $\bb\cdot\z=\sum_{\mu=0}^p\beta_\mu z_\mu$ (with the convention $z_0=1$).  
The outcome likelihood $P_{\bb} (t | \z)$  and the measure $E$ are given by
\begin{eqnarray}
\hspace*{-15mm}
P_{\bb} (t | \z) &=& \Big( \frac{1}{1 + \rm\rme^{-\bb\cdot \z}} \Big)^{\!t}   \Big( \frac{1}{1 + \rm\rme^{\bb\cdot \z}} \Big)^{\!1-t}  
\\
\hspace*{-15mm}
E(\bb^\star , \data) &=& \frac{1}{N} \sum_{i=1}^N \Big\{  
t_i \log \Big( \frac{1 + \rm\rme^{-\bb\cdot \z_i}}{1 + \rm\rme^{-\bb^\star\cdot \z_i}} \Big)  +
(1 - t_i) \log \Big( \frac{1 + \rm\rme^{\bb\cdot \z_i}}{1 + \rm\rme^{\bb^\star\cdot \z_i}} \Big)
\Big\}
\label{eq:overfit}
\end{eqnarray} 
Results are shown in Figure \ref{fig:ML_intuition}. 
When $\zz = 0.025$,  $E$ converges towards zero during the minimization, indicating perfect parameter recovery. As the number of samples in the data set is reduced, giving $\zz=0.25$ and $\zz=2.5$, $E$ converges to increasingly negative values. Since there is no model mismatch (the data were generated from a logistic model), the negative values of $E$ indicate overfitting.  
\vsp

Switching  from maximum likelihood to maximum a posteriori estimators implies adding a penalty term to the likelihood: $D(\hat{P}_{\data} \|  P_{\param} )\to D(\hat{P}_{\data} \|  P_{\param} )  - \log p(\param)$ where $p(\param)$ represents a parameter prior, giving
\begin{eqnarray}
\hspace*{-15mm} E(\param^\star , \data) &\equiv& \underset{\param}{\min} \, \bigg\{ D(\hat{P}_{\data}  \|  P_{\param} ) - \log p(\param) \bigg\}  - \bigg\{ D(\hat{P}_{\data}  \|  P_{\param^\star} )  - \log p(\param^\star) \bigg\} \nonumber  
\\
&=& \underset{\param}{\min}  \bigg\{ \frac{1}{N} \,  \sum_{i=1}^N \log \frac{p(t_i | \z_i, \param^\star) \, p(\param^\star)}{p(t_i | \z_i, \param) \, p(\param)} \bigg\} 
\label{eq:minimize}
\end{eqnarray}
MAP regression is equivalent to minimizing the quantity (\ref{eq:minimize}). 
This minimization should in principle be over all $\param$, but may in practice be constrained to simplify the calculation (see e.g. \cite{fyodorov2018spin,shinzato2018replica,ciliberti2007risk}). 
For generalized linear models, commonly used priors are $p(\bb) \propto\exp[- \eta\sum_{\mu=1}^p |\beta_{\mu}|]$ (giving $L_1$ regularization\footnote{This choice promotes sparsity in the regression coefficient vector $\bb$, which would result in a horizontal line segment passing through the origin in Fig. \ref{fig:slope}. Since our theory aims to predict the slope of the data clouds in Fig. \ref{fig:slope}, we will not pursue $L_1$ regularizers in this paper.},  or `LASSO' regression \cite{tibshirani1996regression}) and 
$p(\bb) \propto \exp[-\eta \sum_{\mu=1}^p \beta_{\mu}^2]$  (giving $L_2$ regularization, or `ridge' regression).   

In the present  paper we generalize the replica analysis of  \cite{coolen2017replica} from ML to MAP inference, upon adding an $L_2$ regularization term to the log-likelihood function.  This term suppresses overfitting effects, and removes the ML phase transition of the Cox model \cite{coolen2017replica} at $\zeta=1$; see e.g. Fig. \ref{fig:slope}. In the presence of an $L_2$ regularizer, correlations between covariates can no longer be transformed away, as was done in \cite{coolen2017replica}, leading to the appearance  in the theory of the population covariance matrix $\A$ of the covariates. Under mild restrictions on the eigenvalue spectrum of this matrix, we show how an accurate theory of overfitting for the regularized Cox proportional hazards model can be developed, in spite of such additional mathematical complications, including for the previously inaccessible regime $\zeta >1$. 
We find, as in \cite{coolen2017replica},  that the replica symmetric version of the theory is sufficient to explain accurately the behaviour of interest. The resulting equations can also be used to predict the amount of regularization needed  for unbiased regression, expressed in term of spectrum of $\A$ and the ratio $\zeta$.

\section{Replica analysis of regularized Cox regression }
\label{ssection:replica}

\subsection{Generalized replica formalism to include priors}

Following \cite{coolen2017replica}, we interpret minimization of (\ref{eq:minimize}) as computing the  ground state energy of a statistical mechanical system with degrees of freedom $\param$ and Hamiltonian $H(\param|\param^\star\!,\data)$, at inverse temperature $\gamma$, where $\data=\{(t_1, \z_1), \ldots, (t_N, \z_N)  \}$ and
\begin{eqnarray}
H(\param|\param^\star\!,\data)&=& \log \prod_{i=1}^N \Big[\frac{p(t_i | \z_i, \param^\star) \, p(\param^\star)}{p(t_i | \z_i, \param) \, p(\param)}\Big]
\end{eqnarray} 
We define the associated free energy, which we average over the disorder (the microscopic realization of $\data$), and can compute the disorder-averaged ground state energy  as the $\gamma\to\infty$ limit of the disorder-averaged energy density $E_{\gamma}(\param^\star)$, where
\begin{eqnarray}
\label{eq:hamiltonian}
E_{\gamma}(\param^\star)
&=&  - \frac{1}{N} \frac{\partial }{\partial \gamma} \bigg\langle \log \int\!\rmd \param~ \prod_{i=1}^N  \bigg[ \frac{p(t_i | \z_i, \param)p(\param)}{p(t_i | \z_i, \param^\star)p(\param^\star)}  \bigg]^{\gamma} \bigg\rangle_{\!\data}  \\  \nonumber
\end{eqnarray}
The replica identity 
 $\langle\log Z \rangle = \lim_{n \to 0} n^{-1} \log \langle Z^n \rangle $ is subsequently used to simplify  the average of the logarithm (see \cite{coolen2017replica} for details and references), giving in the present case
 \begin{eqnarray}
 \hspace*{-20mm}
E_{\gamma}(\param^\star) &=&  -\frac{\partial }{\partial \gamma} \lim\limits_{n \to 0} \frac{1}{Nn} \log  \bigg\langle \bigg\{ \int \!\rmd \param ~\prod_{i=1}^N  \bigg[ \frac{p(t_i | \z_i, \param) p(\param) }{p(t_i | \z_i, \param^\star)p(\param^\star)}  \bigg]^{\gamma} \bigg\}^n \bigg\rangle_{\!\data} 
\nonumber \\
\hspace*{-20mm}
&=&  -  \frac{\partial }{\partial \gamma} \lim\limits_{n \to 0} \frac{1}{Nn} \log    \int\! \Big\{\prod_{\alpha=1}^n\rmd \param^\alpha\Big[ \frac{ p(\param^{\alpha}) }{p(\param^\star)}\Big]^\gamma \Big\}
 \bigg\langle \prod_{i=1}^N  \prod_{\alpha=1}^n \Big[ \frac{p(t_i | \z_i, \param^{\alpha})  ) }{p(t_i | \z_i, \param^\star)}  \Big]^{\gamma}  \bigg\rangle_{\!\data}
 \nonumber \\
\hspace*{-20mm}
 &=&  -  \frac{\partial }{\partial \gamma} \lim\limits_{n \to 0} \frac{1}{Nn} \log    \int \!
\Big\{ \prod_{\alpha=1}^n \rmd\param^\alpha \Big[\frac{ p(\param^{\alpha}) }{p(\param^\star)}  \Big]^{\gamma}  \Big\}
 \nonumber \\
\hspace*{-20mm} && \hspace*{20mm}\times \Big\{
\int\! \rmd\z \rmd t~ p(\z)p(t | \z, \param^\star)  
\prod_{\alpha=1}^n \Big[   \frac{p(t | \z, \param^{\alpha})  ) }{p(t | \z, \param^\star)  }    \Big]^{\gamma}
\Big\}^N 
\label{eq:replica1}
\end{eqnarray}
Equation (\ref{eq:replica1})  is applicable to any parametric model $p(t | \z, \param)$ and any prior $p(\param)$. See also  \cite{barbier2017stochastic} for alternative results on the use of the replica method in statistical inference.  We will now make a specific choice for $p(t | \z, \param)$, and use (\ref{eq:replica1}) to develop a theory for regression and overfitting in regularized Cox models with Gaussian priors. 

\subsection{Application to the regularized Cox proportional hazards model}

Cox's proportional hazards model originally described in \cite{cox1972regression} assumes a parametrization of the form\footnote{The Cox proportional hazards model can be writen in terms of the probability density function for $t$, the survival function and the cumulative hazard rates. Deriviations of the relationships between these functions can be found in e.g. \cite{klein2006survival}}
\begin{eqnarray}
 p(t|\z,\param)&=&\lambda(t)\rm\rme^{\bb\cdot\z-\exp(\bb\cdot\z)\int_0^t\rmd t^\prime ~\lambda(t^\prime)}
 \end{eqnarray}
where the random variable $t \in \mathbb{R}^+$ represents the time-to-event/failure for the sample. Its parameters are the coefficients $\bb \in \mathbb{R}^p$, and a base hazard rate $\lambda(t)$ (a nonnegative function defined for $0\leq t<\infty$). For practical use, the focus is often on the so-called hazard ratios which compare the values of the factors $\exp(\beta_\mu z_\mu)$ for different covariate values. In this case, no assumptions are required for the unknown $\lambda(t)$ beyond $\lambda(t)\geq 0$. For our replica analysis, we make a variational approximation described in Section \ref{sec:va}. Substituting $\param = \{\bb, \lambda \}$ translates (\ref{eq:replica1})  into
\begin{eqnarray}
\hspace*{-15mm} E_{\gamma}(\bb^\star, \lam^\star) &=& - \frac{\partial }{\partial \gamma}  \lim\limits_{n \to 0} \frac{1}{Nn}
\log \int \{\rmd\lam^1\! \ldots \rmd\lam^n\}   \int\!\rmd\bb^1\! \ldots \rmd\bb^n
\Big\{ \prod_{\alpha=1}^n \Big[\frac{ p(\bb^{\alpha}) }{p(\bb^\star)}  \bigg]^{\gamma}  \Big\} 
\nonumber \\
\hspace*{-15mm} &&\times \Big\{
\int\! \rmd\z \rmd t~ p(\z) p(t | \z, \bb^\star\!, \lam^\star)    
\prod_{\alpha=1}^n \Big[   \frac{p(t | \z, \bb^{\alpha}\!, \lam^{\alpha}) ) }{p(t | \z, \bb^\star\!, \lam^\star) } \Big]^{\gamma}
\Big\}^N
\label{eq:energy}
\end{eqnarray}
 Functional integrals are written as $\int\{\rmd\lambda\}$, the true parameters responsible for the data are written as $\{\bb^\star\!, \lambda^\star \}$, and we follow the standard convention for regularized Cox models of only including a prior for the association parameters (equivalently, assuming an improper, or `flat', prior for the base hazard rate). Our $L_2$ prior is $p(\bb)\propto \exp(-p\eta\bb^2)$, and we will find in our analysis that this form indeed gives the appropriate scaling with $p$. 
 To proceed with the analytical treatment, we assume that the covariate vectors $\z_i$ are drawn independently from a population distribution with zero mean and covariance matrix $\A$.  The introduction of regularization means that the regression equations for correlated covariates can no longer be transformed to those corresponding to uncorrelated ones. This leads to a more complex theory than \cite{coolen2017replica}, and ultimately to conditions on the eigenvalue spectrum of $\A$. 

Our analysis is carried out in the regime where both $N,p \rightarrow \infty$ but with fixed ratio $\zz = p/N \sim \mathcal{O}(1)$. To retain non-zero event times, even for $p \rightarrow \infty$, we must rescale the regression coefficients according to $\bb\rightarrow \bb/\sqrt{p}$, resulting in $\bb\cdot \z \sim \mathcal{O}(1)$. Without this rescaling we would have  event time distributions with all weight concentrated on $t\to 0$ and $t\to\infty$. We also replace $\bb^\star$ by $\bb^0$, to allow for more compact notation. Following \cite{coolen2017replica} we next introduce
\begin{eqnarray}
p(\y | \bb^0\!, \ldots, \bb^n) = \int\!\rmd \z~ p(\z) \prod_{\alpha=0}^n \delta \Big[ y^{\alpha} -   \frac{\bb^{\alpha}\! \cdot \z}{\sqrt{p}} \Big]
\end{eqnarray}
where $\y = \{ y^0 , y^1, \ldots, y^n \}\! \in\! \mathbb{R}^{n+1}$. The magnitude of $\rm\rme^{\bb\cdot\z}$ represents the relative risk of failure, compared  to that of an `average' individual (with $\z=\bnull$). Therefore $\y$ can be considered a vector of risk scores. Our energy density then becomes
\begin{eqnarray}
\label{eq:energy2}
\hspace*{-20mm}
E_{\gamma}(\bb^\star, \lam^\star) &=& - \frac{\partial }{\partial \gamma}  \lim\limits_{n \to 0} \frac{1}{Nn}
\log \int \!\{\rmd\lam^1\! \ldots \rmd\lam^n\}   \int \!\rmd\bb^1\! \ldots \rmd\bb^n  \prod_{\alpha=1}^n \Big[\frac{ p(\bb^{\alpha}) }{p(\bb^0)}  \bigg]^{\gamma} 
\nonumber \\
\hspace*{-20mm}
&& \times \Big\{
\int\! \rmd \y ~ p(\y | \bb^0\!, \ldots, \bb^n) \int \!\rmd t~ p(t |y^0\!, \lam^0)
\prod_{\alpha=1}^n \Big[   \frac{p(t |y^{\alpha}\!, \lam^{\alpha}) }{p(t |y^0\!, \lam^0) }    \Big]^{\gamma}
\Big\}^N 
\end{eqnarray}
in which now
\begin{eqnarray}
 p(t|y,\lambda)&=&\lambda(t)\rme^{y-\exp(y)\int_0^t\rmd t^\prime ~\lambda(t^\prime)}
 \end{eqnarray}
 To proceed we assume that $p(\y | \bb^0, \ldots, \bb^n)$ is Gaussian. This holds for any $N$ and $p$ as soon as $p(\z)$ is Gaussian, and for non-Gaussian covariate statistics it will generally hold due to the Central Limit Theorem if the correlations among the covariates  are weak, and  $N$ and $p$ are  large. 
  Since we assumed $\int\!\rmd\z~p(\z)\z=\bnull$, the risk score distribution is now given by
 \begin{eqnarray}
p(\y | \bb^0, \ldots, \bb^n) = \frac{\rm\rme^{-\half \y\cdot \C^{-1}[\{\bb \}] \y}}{\sqrt{(2 \pi)^{n+1}  \det \C[\{\bb \}]}}
\label{eq:y_mvn}
\end{eqnarray}
It is determined in full by the $(n\!+\!1)\! \times\! (n\!+\!1)$ covariance matrix $\C[\{\bb \}]$, with entries
\begin{eqnarray}
C_{\alpha \rho}[\{\bb \}] &=& \int\! \rmd\z~ p(\z) \Big(\frac{\bb^{\alpha}\!\cdot\z}{\sqrt p}\Big)\Big( \frac{\bb^{\rho}\!\cdot\z}{\sqrt p} \Big) 
~= \frac{1}{p} \bb^{\alpha}\!\cdot \A \bb^{\rho}  
\label{eq:correlated}
\end{eqnarray}
The entries of $\A$ are given by $A_{\mu \nu} = \int\!\rmd\z~p(\z) z_{\mu} z_{\nu}$. The $\{C_{\alpha \rho}[\{\bb \}]\}$ measure the similarity between the $p$-dimensional vectors formed by the regression parameters in different replicas. For each replica pair $(\alpha,\rho)$ we use the integral representation of the Dirac delta function, and rescale the conjugate integration parameter by $p$, 
\begin{eqnarray}
\hspace*{-10mm}
1 = \int\!   \rmd C_{\alpha \rho}~\delta \big[ C_{\alpha \rho} \!-\! \frac{1}{p} \bb^{\alpha}\!\cdot \A \bb^{\rho} \big] =
\int \!\frac{\rmd C_{\alpha \rho} \rmd \Chat_{\alpha \rho}}{{2 \pi}/p} \rm\rme^{\rmi p \Chat_{\alpha \rho} (C_{\alpha \rho} - \frac{1}{p} \bb^{\alpha}\!\cdot \A \bb^{\rho} )}
\end{eqnarray}
in order to simplify expression  (\ref{eq:energy2}) to
\begin{eqnarray}
\hspace*{-15mm}
 E_{\gamma}(\bb^\star, \lam^\star) &=& - \frac{\partial }{\partial \gamma}  \lim\limits_{n \to 0} \frac{1}{Nn}
\log \int \!\{\rmd \lam^1\! \ldots \rmd \lam^n\}   \int\! \rmd \C \, \rmd \hat{\C}~\frac{\rm\rme^{\rmi p \, \sum_{\alpha ,\rho=0}^n \Chat_{\alpha \rho} \, C_{\alpha \rho} }}{(2 \pi / p  )^{(n+1)^2}} \nonumber \\
&& \times \Bigg[
\int \! \frac{ \rmd \y~\rm\rme^{-\half \y^T \C^{-1} \y}}{\sqrt{(2 \pi)^{n+1}  \det \C}} \int \! \rmd t~ p(t | y^0\!, \lam^0)
\prod_{\alpha=1}^n \Big[   \frac{p(t | y^{\alpha}\!, \lam^{\alpha})  }{p(t | y^0\!, \lam^0)  }    \Big]^{\gamma}
\Bigg]^N \nonumber \\
&& \times \int\! \rmd\bb^1\! \ldots d\bb^n~  \rm\rme^{-\eta \gamma \sum_{\alpha=1}^n [ (\bb^{\alpha})^2  -  (\bb^{0})^2 ] -\rmi  \sum_{\alpha, \rho=0}^n  \Chat_{\alpha \rho}  \bb^{\alpha} \cdot\A \bb^{\rho }}  
\label{eq:energy3a}
\end{eqnarray}
The quadratic nature of the exponent in the $\bb$ integral, a consequence  of having chosen $L_2$ regularization, allows for a closed form solution. Changing the penalty term to $L_1$ or $L_q$ with $q \!>\! 2$ would significantly complicate the integrals. 

\subsection{Conversion into a saddle point problem}

With a modest amount of foresight we transform $\hat{\C}=-\frac{1}{2}\rmi\D$, and introduce the short-hand $\tbb \equiv \A^{\half} \bb$.
To evaluate the Gaussian $\bb$ integral in \eref{eq:energy3a}  we define the $np\times np$ matrix $\bXi $ and the $np$-dimensional vector $\bxi$, with entries
\begin{eqnarray}
\Xi_{\alpha\mu;\beta\nu}= 2 \eta \gamma \delta_{\alpha \beta} (\A^{-1})_{\mu \nu} +   \delta_{\mu \nu}D_{\alpha \beta},~~~~~~ \xi_{\mu}^{\alpha}= -D_{0 \alpha}  \tb_{\mu }^{0}
\label{eq:define_xi}
\end{eqnarray}
With these definitions we may write the Gaussian integral  in (\ref{eq:energy3a}) as
\begin{eqnarray}
\hspace*{-0mm} 
&&
\fl\int  \Big( \prod_{\alpha=1}^n \rmd \tbb^{\alpha} \rme^{-\eta \gamma \tbb^{\alpha}\cdot \A^{-1} \tbb^{\alpha}}\Big)
\rme^{ -\frac{1}{2}  \sum_{\alpha,\rho=1}^n  D_{\alpha \rho}  \tbb^{\alpha} \cdot \tbb^{\rho }  - \sum_{\rho=1}^n D_{0 \rho}  \tbb^{0} \cdot \tbb^{\rho } }  
\nonumber  \\  
\hspace*{-2mm} 
&=& 
 \rme^{\half \bxi\cdot \bXi^{-1} \bxi} \! \int\! \rmd \tbb ~ \rme^{ -\half (\tbb - \bXi^{-1} \bxi)\cdot \bXi (\tbb - \bXi^{-1} \bxi)  }  
= 
 \frac{(2\pi)^{\frac{np}{2}}}{\sqrt{\det\bXi}} \rme^{\half \bxi\cdot \bXi^{-1} \bxi} 
\label{eq:gaussian}
\end{eqnarray}
Let $\{a_\mu\}$ and $\{b_\alpha\}$  denote the eigenvalues of $\A$ and $\D$, respectively.  
The two terms $\PP$ and $\QQ$ of the matrix $\bXi$, with components 
$P_{\alpha\mu,\beta\nu}=2 \eta \gamma \delta_{\alpha \beta} (\A^{-1})_{\mu \nu} $ and $Q_{\alpha\mu,\beta\nu}= \delta_{\mu \nu} D_{\alpha \beta}$,  clearly commute.
The complete set of eigenvectors  of $\btXi$ can therefore be written as $\{\hat{\uu}^{\mu\alpha}\}$, with components 
$\hat{u}^{\mu\alpha}_{\nu\rho}=u^\alpha_\rho v^\mu_{\nu}$, and where $\sum_{\rho\leq n} D_{\lambda\rho}u_\rho^\alpha=b_\alpha u_\rho^\lambda$ and 
$\sum_{\nu\leq p}A_{\lambda\nu} v^{\mu}_{\nu}=a_\mu v^{\mu}_{\lambda}$, and where both are normalised according to $\sum_{\rho\leq n}(u^\alpha_\rho)^2=\sum_{\nu\leq p} (v^\mu_{\nu})^2=1$. 
The eigenvalues of $\btXi$ are then $\xi_{\mu\alpha} =2 \eta \gamma/a_\mu + b_\alpha$, and 
\begin{eqnarray}
\hspace*{-15mm}
\det \btXi = \prod_{\mu=1}^p \prod_{\alpha=1}^n \big( \frac{2 \eta 
\gamma}{a_\mu} \!+\! b_\alpha  \big),~~~~~~
(\bXi^{-1})_{\alpha\mu,\alpha^\prime\mu^\prime}= \sum_{\beta=1}^n \sum_{\nu=1}^p 
\frac{u^\beta_\alpha v^\nu_{\mu} u^\beta_{\alpha^\prime} v^\nu_{\mu^\prime}}{2 \eta \gamma/a_\nu + b_\beta}
\end{eqnarray}
  Hence the integral (\ref{eq:gaussian}) can be written as
\begin{eqnarray}
\hspace*{-15mm}
\frac{(2\pi)^{\frac{np}{2}}}{\sqrt{{\rm det}\bXi}}~ \rme^{\half \bxi\cdot \bXi^{-1} \bxi} &=&
\rme^{\frac{1}{2}np\log(2\pi)-\frac{1}{2}np \big\langle\log ( 2 \eta 
\gamma/a +b )\big\rangle+
\frac{1}{2} np\big\langle (\bxi\cdot\hat{\uu})^2
(2 \eta \gamma/a + b)^{-1}\big\rangle}
\label{eq:gaussian_done}
\end{eqnarray}
where the averages in the exponents are over the eigenvalues and orthonormal eigenvectors of $\bXi$, i.e. 
$\langle f(a,b,\hat{\uu})\rangle=(np)^{-1}\sum_{\mu=1}^p\sum_{\alpha=1}^n f(a_\mu,b_\alpha,\hat{\uu}^{\mu\alpha})$.
Since $p=\zeta N$ with $\zeta\!>\!0$, the integrals over $\C$, $\hat{\C}$ and the base hazard rates in (\ref{eq:energy3a}) can for $N\to\infty$ be evaluated by steepest descent, 
provided the limits $n\!\to\! 0$ and $N\!\to\! \infty$ commute. 
Expression (\ref{eq:gaussian_done}) then enables us to write the result as
\begin{eqnarray}
\lim\limits_{N \to \infty} E_{\gamma}(\bb^\star, \lam^\star) &=& \frac{\partial }{\partial \gamma}\lim\limits_{n \to 0} \frac{1}{n} \mbox{extr}\, \Psi(\C, \D, \lam^1 \ldots \lam^n)    
\label{eq:energy3d}
\end{eqnarray}
in which
\begin{eqnarray}
\hspace*{-20mm}
 \Psi (\C, \D, \lam^1\! \ldots \lam^n) &=&  -\frac{1}{2} \zeta \, \bigg[\sum_{\alpha, \rho=0}^n D_{\alpha \rho} C_{\alpha \rho} - 
 \frac{1}{p}D_{00} (\tbb^0)^2  \bigg]  
 +\half (n\!+\!1\!-\!n\zeta) \log(2 \pi) 
 \nonumber\\
  \hspace*{-20mm}
&&\hspace*{-10mm}
 + \half \log \det \C  - n \eta \zz \gamma S^2  
 + \half n\zz  \Big\langle\!  \log  \Big( \frac{2 \eta \gamma}{a}\! +\! b  \Big)   \Big\rangle
  - \half n\zz  \Big\langle \frac{(\bxi\cdot\hat{\uu})^2}{2 \eta \gamma/a\! +\! b} \Big\rangle
 \nonumber  \\
 \hspace*{-20mm}
&& \hspace*{-10mm}
- \log  \int \!\rmd \y ~\rme^{-\half \y\cdot \C^{-1} \y} \! \int\!\rmd t~ p(t |y^0\!, \lam^0)
\prod_{\alpha=1}^n \Big[   \frac{p(t | y^{\alpha}\!, \lam^{\alpha})  }{p(t | y^0\!, \lam^0)  }    \Big]^{\gamma}
  \label{eq:energy3c}
\end{eqnarray}
where we have defined $S^2=\lim_{p\to\infty}p^{-1}(\bb^0)^2$.
Differentiating $\Psi$ with respect to $D_{00}$ removes $D_{00}$ from the problem, and gives $C_{00} = p^{-1}\bb^0\cdot \A \bb^0 \equiv \tS^2 $.  

\section{Replica symmetric theory}

\subsection{Replica symmetric saddle points}

To proceed, we make the replica symmetric ansatz, which implies assuming ergodicity of the stochastic regression process, and  translates into invariance of all order parameters under all permutations of the replicas $\{1,\ldots,n\}$. Now, for all $1\leq \alpha, \rho \leq n$:
\begin{eqnarray}
\hspace*{-10mm}
\lam^{\alpha}(t) = \lam(t),~~~~~~
\begin{array}{l} 
C_{0 \alpha} = c_0
\\[1mm]
D_{0\alpha}=d_0
\end{array},~~~~~~
\begin{array}{ll}
C_{\alpha \rho} & = C \delta_{\alpha \rho} + c (1 - \delta_{\alpha \rho} ) 
\\[1mm]
D_{\alpha \rho} & = D \delta_{\alpha \rho} + d (1 - \delta_{\alpha \rho} ) 
\end{array}
\end{eqnarray}
Both $\C$ and $\D$ are positive definite, so $C>c$ and $D>d$.  We may now write
\begin{eqnarray}
&&
\hspace*{-10mm}
 \C = \pmatrix{%
C_{00} & c_0 &\ldots & \ldots & c_0 \cr
c_0 & C & c & \ldots & c \cr
\vdots & c & C & \ldots & c \cr
\vdots & \vdots & \vdots & \ddots & \vdots \cr
c_0 & c & c & \ldots & C \cr
},~~~~~~
\C^{-1} = \pmatrix{%
B_{00} & b_0 &\ldots & \ldots & b_0 \cr
b_0 & B & b & \ldots & b \cr
\vdots & b & B & \ldots & b \cr \nonumber
\vdots & \vdots & \vdots & \ddots & \vdots \cr
b_0 & b & b & \ldots & B \cr
}
\label{eq:Cmatrix}
\end{eqnarray}
The eigenvalues and eigenvectors of $\C$, $\C^{-1}$ and $\D$ are found in  \cite{coolen2017replica}.
$\C$ has two nondegenerate eigenvalues $\lambda_{\pm}$ with 
 $\lambda_+ \lambda_- =  [C+(n\!-\!1)c ] C_{00} - nc_0^2$,  and a further $n\!-\!1$ fold degenerate eigenvalue $\lambda_0=C-c$. Hence 
 \begin{eqnarray}
\label{eq:logdetC}
\log \det \C &=&   \log \Big([C+(n\!-\!1)c ] C_{00} - nc_0^2\Big)+(n\!-\!1)\log(C\!-\!c)
\nonumber \\
&=&   \log C_{00} + n \log (C\!-\!c) + \frac{n\big(c\! -\! c_0^2 / C_{00}  \big)}{C-c} + \mathcal{O}(n^2)   \\  \nonumber
\end{eqnarray}
The entries of $\C^{-1}$ are found to be 
\begin{eqnarray}
&& 
\hspace*{-15mm}  B_{00} =  \frac{C + (n-1)c}{C_{00} [C + (n-1)c] - nc_0^2},~~~~~~
 b_0= - \frac{c_0}{C_{00} [C + (n-1)c] - nc_0^2}
  \\  
&& 
\hspace*{-15mm}  B = b + \frac{1}{C-c},~~~~~~\hspace*{15mm}
 b =  \frac{c_0^2 - c C_{00}}{(C_{00} [C + (n-1)c] - nc_0^2)(C-c)}
 \label{eq:Cinv}
\end{eqnarray}
Hence
\begin{eqnarray}
\label{eq:quadratic} 
\hspace*{-15mm}
\y\cdot \C^{-1} \y &=&  B_{00} (y^0)^2 + (B\!-\!b) \sum\limits_{\alpha=1}^n (y^{\alpha})^2 +  b \Big( \sum\limits_{\alpha=1}^n y^{\alpha} \Big)^2  + 2 b_0 y^0 \sum\limits_{\alpha=1}^n y^{\alpha}  \\  \nonumber
\end{eqnarray}
Next we turn to terms in (\ref{eq:energy3c}) that involve the spectrum of $\D$. This  matrix has 
one eigenvalue $D\!+\!(n\!-\!1)d$ with eigenvector $\vv = (1,\ldots, 1)$, and the $n\!-\!1$ fold degenerate  eigenvalue $D-d$ with eigenspace $\sum_{i=1}^n \vv_i = 0$. Hence
\begin{eqnarray}
\hspace*{-15mm}
 \Big\langle\!  \log  \Big( \frac{2 \eta \gamma}{a}\! +\! b  \Big)   \Big\rangle
 &=&
 \frac{1}{np}\sum_{\mu=1}^p \Big[
 (n\!-\!1) \log  \Big( \frac{2 \eta \gamma}{a}\! +\! D\!-\!d  \Big) 
 + \log  \Big( \frac{2 \eta \gamma}{a}\! +\! D\!-\!d \!+\!nd \Big) \Big]
 \nonumber
 \\
 &=& 
\Big\langle \log  \Big( \frac{2 \eta \gamma}{a}\! +\! D\!-\!d  \Big) \Big\rangle
 + \Big\langle\frac{da}{2 \eta \gamma\! +\! (D\!-\!d)a} \Big\rangle +O(n)
 \label{eq:RSterm1}
 \end{eqnarray}
Similarly, using the RS form of $\xi_\mu^\alpha=-d_0(\A^{\frac{1}{2}}\bb^0)_\mu$, we may write 
\begin{eqnarray}
\hspace*{-5mm}
\Big\langle \frac{(\bxi\cdot\hat{\uu})^2}{2 \eta \gamma/a\! +\! b} \Big\rangle &=&
\frac{1}{np}\sum_{\mu=1}^p \Big(\sum_{\nu=1}^p\sum_{\rho=1}^n (\A^{\frac{1}{2}}\bb^0)_\nu v_\nu^\mu\frac{1}{\sqrt{n}}\Big)^2\frac{d_0^2}{2 \eta \gamma/a_\mu\! +\! D\!+\!(n\!-\!1)d}
\nonumber
\\
\hspace*{-5mm}
&=&
\frac{1}{p}\sum_{\mu=1}^p (\bb^0\!\cdot\! \vv^\mu)^2
\frac{d_0^2 a_\mu}{2 \eta \gamma/a_\mu\! +\! D\!-\!d}+O(n)
\nonumber
\\
\hspace*{-5mm}
&=&
d_0^2 ~
\Big\langle 
\frac{a^2 (\bb^0\!\cdot\! \vv)^2}{2 \eta \gamma\! +\! (D\!-\!d)a}\Big\rangle+O(n)
 \label{eq:RSterm2}
\end{eqnarray}
The averages in (\ref{eq:RSterm1},\ref{eq:RSterm2}) are now over the joint distribution of eigenvalues and eigenvectors of $\A$ only. Inserting the above RS expressions into (\ref{eq:energy3c}), and using $C_{00}=\tilde{S}^2$, then gives us, with the short-hand ${\rm D}z=(2\pi)^{-\frac{1}{2}}\rme^{-\frac{1}{2}z^2}\rmd z$, 
\begin{eqnarray}
\hspace*{-20mm}
\frac{1}{n} \Psi_{\rm RS} (\ldots) &=&  -\frac{1}{2} \zeta (
 2d_0c_0+DC-dc)
 +\half (1\!-\!\zeta) \log(2 \pi)  -  \eta \zz \gamma S^2  
 +O(n)
 \nonumber\\
  \hspace*{-20mm}
&&
 + \half 
 \Big[
 \log (C\!-\!c) + \frac{c\! -\! c_0^2 / C_{00}}{C\!-\!c} 
 \Big]
   - \half \zz 
  d_0^2 ~
\Big\langle 
\frac{a^2 (\bb^0\!\cdot\! \vv)^2}{2 \eta \gamma\! +\! (D\!-\!d)a}\Big\rangle
 \nonumber
 \\
 \hspace*{-20mm}
 &&
 + \half \zz 
 \Big\langle \log  \Big( \frac{2 \eta \gamma}{a}\! +\! D\!-\!d  \Big) \Big\rangle
 +  \half \zz \Big\langle\frac{da}{2 \eta \gamma\! +\! (D\!-\!d)a} \Big\rangle 
+\frac{1}{n}\log\tilde{S} 
  \nonumber  \\
 \hspace*{-20mm}
&& 
 -\frac{1}{n}
 \log \int\!{\rm D}z \int\!\frac{\rmd  y_0}{\sqrt{2\pi}} \rme^{-\half 
  B_{00} y_0^2 } \int\!\rmd t~ p(t |y_0, \lam^0)
  \nonumber
  \\
  \hspace*{-20mm}
 &&\times
 \Big[
  \int \!\rmd y~
  \rme^{-\half  (B-b)  y^2 + y(\rmi  \sqrt{b}   -b_0 y_0)} 
   \frac{p^\gamma(t | y, \lam)  }{p^\gamma(t | y_0, \lam^0)  }  \Big]^n
   \nonumber
   \\
   \hspace*{-20mm}
   &=&
     -\frac{1}{2} \zeta (
 2d_0c_0+DC-dc)
 +\half (1\!-\!\zeta) \log(2 \pi)  -  \eta \zz \gamma S^2  
 +O(n)
 \nonumber\\
  \hspace*{-20mm}
&&
 + \half 
 \Big[
 \log (C\!-\!c) + \frac{c\! -\! c_0^2 /\tilde{S}^2}{C\!-\!c} 
 \Big]
   - \half \zz 
  d_0^2 ~
\Big\langle 
\frac{a^2 (\bb^0\!\cdot\! \vv)^2}{2 \eta \gamma\! +\! (D\!-\!d)a}\Big\rangle
 \nonumber
 \\
 \hspace*{-20mm}
 &&
 + \half \zz 
 \Big\langle \log  \Big( \frac{2 \eta \gamma}{a}\! +\! D\!-\!d  \Big) \Big\rangle
 +  \half \zz \Big\langle\frac{da}{2 \eta \gamma\! +\! (D\!-\!d)a} \Big\rangle 
+\frac{1}{2n}\log(\tilde{S}^2  B_{00})
  \nonumber  \\
 \hspace*{-20mm}
&& 
 -\frac{1}{n}
 \log \int\!{\rm D}z {\rm D}y_0\int\!\rmd t~ p(t |y_0/\sqrt{B_{00}}, \lam^0)
  \nonumber
  \\
  \hspace*{-20mm}
 &&~~~\times
 \Big[
  \int \!\rmd y~
  \rme^{-\half  (B-b)  y^2 + y(\rmi  \sqrt{b}   -b_0 y_0/\sqrt{B_{00}})} 
   \frac{p^\gamma(t | y, \lam)  }{p^\gamma(t | y_0/\sqrt{B_{00}}, \lam^0)  }  \Big]^n
\end{eqnarray}
We note that 
\begin{eqnarray}
&&\hspace*{-5mm} 
B_{00}^{-1}=\tilde{S}^2-nc_0^2/(C\!-\!c)+O(n^2),~~~~~~
B-b= 1/(C\!-\!c)
\\
&&\hspace*{-5mm} 
b_0=-c_0/\tilde{S}^2(C\!-\!c)+O(n),~~~~~~~~~~~~~~
b= \frac{c_0^2 - c\tilde{S}^2}{\tilde{S}^2 (C\!-\!c)^2}+O(n)
\end{eqnarray}
and these identities enable us, after some simple rearrangements, to write the limit  $\Psi_{\rm RS}(\ldots)=\lim_{n\to 0}n^{-1}\Psi_{\rm RS}(\ldots)$  in the much simpler form
\begin{eqnarray}
\hspace*{-20mm}
\Psi_{\rm RS}(\ldots)&=&
    -\frac{1}{2} \zeta\Bigg\{
 2d_0c_0+DC-dc
+ \log(2 \pi) +2\eta \gamma S^2  
 \nonumber\\
  \hspace*{-20mm}
&&
 +
  d_0^2 ~
\Big\langle 
\frac{a^2 (\bb^0\!\cdot\! \vv)^2}{2 \eta \gamma\! +\! (D\!-\!d)a}\Big\rangle
-
 \Big\langle \log  \Big( \frac{2 \eta \gamma}{a}\! +\! D\!-\!d  \Big) \Big\rangle
- \Big\langle\frac{da}{2 \eta \gamma\! +\! (D\!-\!d)a} \Big\rangle 
\Bigg\}
  \nonumber  \\
 \hspace*{-20mm}
&& 
\hspace*{-13mm}
 -\! \int\!{\rm D}z {\rm D}y_0\!\int\!\rmd t~ p(t |\tilde{S}y_0, \lam^0)
\log\!
  \int \!{\rm D} y~
   \frac{p^\gamma(t | y\sqrt{C\!-\!c}+\!z(c\!-\!c_0^2/\tilde{S}^2)^{\frac{1}{2}} \!+\! y_0c_0/\tilde{S}, \lam)  }{p^\gamma(t | \tilde{S}y_0, \lam^0)  }  
   \nonumber
   \\[-1mm]
   \hspace*{-20mm}&&
     \label{eq:Psi_RS_before_uvw}
\end{eqnarray}

\subsection{Simplification of the theory and interpretation of order parameters}

Expression    (\ref{eq:Psi_RS_before_uvw}) can readily be extremized over $d_0$, which removes a further order parameter from our theory, and we carry out a suitable transformation of the remaining order parameters, 
\begin{eqnarray}
\hspace*{-15mm}
&& u= \sqrt{C\!-\!c}, ~~~~ v =\sqrt{c\! -\! (c_0/\tS)^2},~~~~ w = c_0/\tS, ~~~~f = d,~~~~ g = D\!-\!d  
\label{eq:transform}
\end{eqnarray}
with $u,v,w\in[0,\infty)$ and with the inverse transformations
\begin{equation}
c_0 = \tS w, \hsp \hsp c = v^2 \!+\! w^2, \hsp \hsp C = u^2 \!+\!v^2\! +\! w^2
\end{equation}
These steps result after some simple rearrangements, and upon removing the term in $\Psi_{\rm RS}$ that will vanish upon differentiation with respect to $\gamma$, in 
\begin{eqnarray}
\lim\limits_{N \to \infty} E_{\gamma}(\bb^\star, \lam^\star) &=& \frac{\partial }{\partial \gamma}\mbox{extr}_{u,v,w,f,g,\lambda} \Psi_{\rm RS}(u,v,w,f,g, \lam)    
\end{eqnarray}
in which
\begin{eqnarray}
\hspace*{-20mm}
\Psi_{\rm RS}(\ldots)   &=&
    -\frac{1}{2} \zeta
    (g\!+\!f)u^2 -\frac{1}{2} \zeta
 g(v^2\!+\!w^2)
-\zeta\eta \gamma S^2  
 \nonumber\\
  \hspace*{-20mm}
&&
\hspace*{0mm}
+\frac{1}{2}\zeta\Bigg\{
 \tilde{S}^2w^2~  \Big\langle 
\frac{a^2 (\bb^0\!\cdot\! \vv)^2}{2 \eta \gamma\! +\! ga}\Big\rangle^{\!-1}\!
+
 \Big\langle\! \log  \Big( \frac{2 \eta \gamma\!+\!ga}{a} \Big) \Big\rangle
+ f~\Big\langle\frac{a}{2 \eta \gamma\! +\! ga} \Big\rangle 
\Bigg\}
  \nonumber  \\
 \hspace*{-20mm}
&& 
 -\! \int\!{\rm D}z {\rm D}y_0\!\int\!\rmd t~ p(t |\tilde{S}y_0, \lam^0)
\log\!
  \int \!{\rm D} y~
   \frac{p^\gamma(t | uy \!+\! wy_0\!+\!vz, \lam)  }{p^\gamma(t | \tilde{S}y_0, \lam^0)  }  
\end{eqnarray}
In principle we could also extremise over $f$, leading to a simple expression with which to remove not just $f$ but also either $u$ or $g$. The true association parameters $\bb^0$ are seen to enter the asymptotic theory only in two places: 
in $\tilde{S}^2=\lim_{p\to\infty}p^{-1}\bb^0\!\cdot\A\bb^0$ and in $\bra a^2(\bb^0\!\cdot\vv)^2/(2\eta\gamma+ga)\ket$. Both are quadratic functions of $\bb^0$. In \ref{app:self_averaging} we show that, if the true associations $\{\beta_\mu^0\}$ are drawn randomly and independently from a zero-average distribution, and under mild conditions on the spectrum $\varrho(a)$ of the covariate correlation matrix $\A$, both terms will be self-averaging with respect to the realization of $\bb^0$. 
Consequently,  with $S^2=\lim_{p\to\infty}p^{-1}(\bb^0)^2$ we may then write
\begin{eqnarray}
\tilde{S}^2= S^2\bra a\ket,~~~~~~
\Big\bra \frac{a^2(\bb^0\!\cdot\vv)^2}{2\eta\gamma\!+\!ga}\Big\ket= 
\bra \frac{S^2 a^2}{2\eta\gamma\!+\!ga}\ket
\end{eqnarray}
(where we used the fact that the eigenvectors $\vv$ of $\A$ were defined to be normalized). Our replica symmetric theory thereby becomes
\begin{eqnarray}
\hspace*{-20mm}
\lim\limits_{N \to \infty} E_{\gamma}(\bb^0, \lam^0) &=&
\int\! {\rm D}y_0\!\int\!\rmd t~ p(t |S\bra a\ket^{\frac{1}{2}}y_0, \lam^0)
\log p(t | S\bra a\ket^{\frac{1}{2}}y_0, \lam^0)  
-\zeta\eta  S^2  
 \nonumber\\
  \hspace*{-20mm}
&&
\hspace*{-21mm}
+\eta\zeta\Bigg\{
w^2\bra a\ket  
 \Big\langle \frac{a^2}{2 \eta \gamma\! +\! ga}\Big\rangle^{\!-2}
 \Big\langle  \frac{a^2}{(2 \eta \gamma\! +\! ga)^2}\Big\rangle
+
 \Big\langle\! \frac{1}{2 \eta \gamma\!+\!ga} \Big\rangle- f~\Big\langle\frac{a}{(2 \eta \gamma\! +\! ga)^2} \Big\rangle 
\Bigg\}
  \nonumber  \\
 \hspace*{-20mm}
&& 
\hspace*{-33mm}
 -\! \int\!{\rm D}z {\rm D}y_0\!\int\!\rmd t~ p(t |S\bra a\ket^{\frac{1}{2}}y_0, \lam^0)
\frac{  \int \!{\rm D} y~
p^\gamma(t | uy \!+\! wy_0\!+\!vz, \lam)\log p(t | uy \!+\! wy_0\!+\!vz, \lam) }
{  \int \!{\rm D} y~
p^\gamma(t | uy \!+\! wy_0\!+\!vz, \lam) }
\nonumber
\\
\hspace*{-20mm}&&
\label{eq:Evalue_before_scaling}
\end{eqnarray}
The scalar order parameters $(u,v,w,f,g\}$ and the function $\lambda(t)$ are computed by extremization of the following function, from which we removed any constant terms:
\begin{eqnarray}
\hspace*{-20mm}
\Psi_{\rm RS}(\ldots)   &=&
    -\frac{1}{2} \zeta
    (g\!+\!f)u^2 -\frac{1}{2} \zeta
 g(v^2\!+\!w^2)
 \nonumber\\
  \hspace*{-20mm}
&&
\hspace*{0mm}
+\frac{1}{2}\zeta\Bigg\{
 w^2\bra a\ket  \Big\langle 
\frac{a^2}{2 \eta \gamma\! +\! ga}\Big\rangle^{\!-1}\!
+
 \Big\langle\! \log (2 \eta \gamma\!+\!ga) \Big\rangle
+ f~\Big\langle\frac{a}{2 \eta \gamma\! +\! ga} \Big\rangle 
\Bigg\}
  \nonumber  \\
 \hspace*{-20mm}
&& \hspace*{-5mm}
 -\! \int\!{\rm D}z {\rm D}y_0\!\int\!\rmd t~ p(t |S\bra a\ket^{\frac{1}{2}}y_0, \lam^0)
\log\!
  \int \!{\rm D} y~
 p^\gamma(t | uy \!+\! wy_0\!+\!vz, \lam)  
 \label{eq:RS_Psi_before_scaling}
\end{eqnarray}
\vsp

The physical meaning of the RS order parameters can be inferred by adapting the route followed in \cite{coolen2017replica}. Upon defining averages over the stochastic MAP minimisation process as $\langle \ldots\rangle$ and those over the realisations of the data set as $\langle \ldots\rangle_D$, this results in
\begin{eqnarray}
&&
\hspace*{-18mm}
C=\lim_{p\to\infty}\frac{1}{p}\bra\bra \bb\!\cdot\!\A\bb\ket\ket_{\cal D},~~~~
c=\lim_{p\to\infty}\frac{1}{p}\bra\bra \bb\ket\!\cdot\!\A \bra \bb\ket\ket_{\cal D},~~~~
c_0=\lim_{p\to\infty}\frac{1}{p}\bb^0\!\cdot\! \A\bra\bra \bb\ket\ket_{\cal D}
\nonumber
\\&&
\label{eq:physical_meaning}
\end{eqnarray}
These order parameters can be used to predict the slope and width of the association parameter cloud in Figure \ref{fig:slope}. A plausible model for this cloud is $\langle \bb \rangle = \kappa \bb^0 + \bomega$, in which $\bomega$ denotes a zero-average inference noise contribution that depends on the data set ${\cal D}$. The entries $\Omega_{\mu\nu}=\langle \omega_{\mu} \omega_{\nu} \rangle_{\cal D} $ of the $p\times p$ noise covariance matrix $\bOmega$  have the same dimension as those of $\A^{-1}$, which prompt us to postulate that $\bOmega=\sigma^2\A^{-1}$. Inserting the above expression for $\bra \bb\ket$ into (\ref{eq:physical_meaning}) leads to the following identities:
\begin{eqnarray}
&& c= 
\kappa^2\tilde{S}^2
+\sigma^2,~~~~~~
c_0=\kappa \tilde{S}^2
\\
&& u^2=\lim_{p\to\infty}\frac{1}{p}\sum_{\mu\nu=1}^p A_{\mu\nu}\bra
\large\bra \beta_\mu\beta_\nu\ket\!-\!\bra \beta_\mu\ket\bra \beta_\nu\ket\large\ket_{\cal D}
\end{eqnarray}
  Using the transformations  \eref{eq:transform} we then obtain the following simple expressions for the two dominant characteristics $\kappa$ and $\sigma$ of the simulation data clouds: 
 \begin{eqnarray}
 \kappa=w/\tilde{S},~~~~~~\sigma=v
 \end{eqnarray} 
 
\subsection{Scaling of order parameters with $\gamma$ }

We will only be interested in the limit $\gamma\to \infty$, where the stochastic process becomes deterministic MAP inference. Following \cite{coolen2017replica},  and with a modest amount of foresight regarding the behaviour of the new order parameters that did not feature in \cite{coolen2017replica}, we make the following ansatz for the scaling with $\gamma$ of the scalar order parameters:
\begin{eqnarray}
u=\tilde{u}/\sqrt{\gamma},~~~~~v,w=O(1),~~~~~g=\tilde{g}\gamma,~~~~~f=\tilde{f}\gamma^2
\end{eqnarray}
 Insertion into (\ref{eq:RS_Psi_before_scaling}), followed by taking the limit $\gamma\to\infty$, gives
 \begin{eqnarray}
\hspace*{-20mm}
\lim_{\gamma\to\infty}\frac{1}{\gamma}\Psi_{\rm RS}(\ldots)   &=&
 \frac{1}{2}\zeta\Bigg\{
 w^2\bra a\ket  \Big\langle 
\frac{a^2}{2 \eta\! +\! \tilde{g}a}\Big\rangle^{\!-1}\!
+\tilde{f}\Big[\Big\langle\frac{a}{2 \eta\! +\! \tilde{g}a} \Big\rangle 
   -
\tilde{u}^2\Big] -
 \tilde{g}(v^2\!+\!w^2)
\Bigg\}
  \nonumber  \\
 \hspace*{-20mm}
&& \hspace*{-21mm}
-\int\!{\rm D}z {\rm D}y_0\!\int\!\rmd t~ p(t |S\bra a\ket^{\frac{1}{2}}y_0, \lam^0)
 \lim_{\gamma\to\infty}\frac{1}{\gamma}
\log\!
  \int \!{\rm d} y~\rme^{\gamma\big[\log 
 p(t | \tilde{u}y+wy_0+vz, \lam)-\frac{1}{2}y^2\big]}  
 \nonumber
 \\
 \hspace*{-20mm}
 &=&
 \frac{1}{2}\zeta\Bigg\{
 w^2\bra a\ket  \Big\langle 
\frac{a^2}{2 \eta\! +\! \tilde{g}a}\Big\rangle^{\!-1}\!
+\tilde{f}\Big[\Big\langle\frac{a}{2 \eta\! +\! \tilde{g}a} \Big\rangle 
   -
\tilde{u}^2\Big] -
 \tilde{g}(v^2\!+\!w^2)
\Bigg\}
  \nonumber  \\
 \hspace*{-20mm}
&& \hspace*{-15mm}
-\int\!{\rm D}z {\rm D}y_0\!\int\!\rmd t~ p(t |S\bra a\ket^{\frac{1}{2}}y_0, \lam^0) {\rm max}_{y}
\Big[\log 
 p(t | \tilde{u}y\!+\!wy_0\!+\!vz, \lam)-\frac{1}{2}y^2\Big]
 \nonumber
 \\
 \hspace*{-20mm}&&
\end{eqnarray}
The maximization over $y$ proceeds as in \cite{coolen2017replica}, giving 
\begin{eqnarray}
\hspace*{-17mm}
 {\rm argmax}_{y}
\Big[\log 
 p(t | \tilde{u}y\!+\!wy_0\!+\!vz, \lam)\!-\!\frac{1}{2}y^2\Big]&=& \tilde{u}-\frac{1}{\tilde{u}}W\Big(\tilde{u}^2\rme^{\tilde{u}^2+wy_0+vz}\Lambda(t)\Big)
 \\
 \hspace*{-12mm}
  {\rm max}_{y}
\Big[\log 
 p(t | \tilde{u}y\!+\!wy_0\!+\!vz, \lam)\!-\!\frac{1}{2}y^2\Big]
&=& \frac{1}{2}(\tilde{u}^2\!+\!\tilde{u}^{-2})+wy_0+vz
 \nonumber
\\
\hspace*{-17mm}&&\hspace*{-25mm}+\log\lambda(t)-\frac{1}{2\tilde{u}^2}\Big[
W\Big(\tilde{u}^2\rme^{\tilde{u}^2+wy_0+vz}\Lambda(t)\Big)\!+\!1\Big]^2
\end{eqnarray}
in which $W(x)$ denotes Lambert's $W$-function, i.e. the inverse of $f(x)=x\rme^x$. This then results in 
\begin{eqnarray}
\hspace*{-20mm}
\lim_{\gamma\to\infty}\frac{1}{\gamma}\Psi_{\rm RS}(\ldots)   &=&
  \frac{1}{2}\zeta\Big[
 w^2\bra a\ket  \Big\langle 
\frac{a^2}{2 \eta\! +\! \tilde{g}a}\Big\rangle^{\!-1}\!
+\tilde{f}\Big[\Big\langle\frac{a}{2 \eta\! +\! \tilde{g}a} \Big\rangle 
   -
\tilde{u}^2\Big] -
 \tilde{g}(v^2\!+\!w^2)
\Big]
\nonumber
\\
\hspace*{-20mm}
&& \hspace*{-5mm}
+\frac{1}{2\tilde{u}^2}\int\!{\rm D}z {\rm D}y_0\!\int\!\rmd t~ p(t |S\bra a\ket^{\frac{1}{2}}y_0, \lam^0)
\Big[
W\Big(\tilde{u}^2\rme^{\tilde{u}^2+wy_0+vz}\Lambda(t)\Big)\!+\!1\Big]^2
  \nonumber  \\
 \hspace*{-20mm}
&&
-\frac{1}{2}(\tilde{u}^2\!+\!\tilde{u}^{-2})
-\int\!{\rm D}y_0\!\int\!\rmd t~ p(t |S\bra a\ket^{\frac{1}{2}}y_0, \lam^0)
\log\lambda(t)
\label{eq:Psi_gamma_infty}
\end{eqnarray}
Similarly, working out (\ref{eq:Evalue_before_scaling}) in the limit $\gamma\to\infty$ gives
\begin{eqnarray}
\hspace*{-20mm}
\lim\limits_{N \to \infty} E_{\infty}(\bb^0, \lam^0) &=&
\int\! {\rm D}y_0\!\int\!\rmd t~ p(t |S\bra a\ket^{\frac{1}{2}}y_0, \lam^0)
\Big[\log p(t | S\bra a\ket^{\frac{1}{2}}y_0, \lam^0)  
-\log\lambda(t)\Big]
 \nonumber\\
  \hspace*{-20mm}
&&
\hspace*{-10mm}
-\tilde{u}^2
-\zeta\eta  S^2  
+\eta\zeta\Big[
w^2\bra a\ket  
 \Big\langle \frac{a^2}{2 \eta\! +\! \tilde{g}a}\Big\rangle^{\!-2}
 \Big\langle  \frac{a^2}{(2 \eta \! +\! \tilde{g}a)^2}\Big\rangle
- \tilde{f}\Big\langle\frac{a}{(2 \eta\! +\! \tilde{g}a)^2} \Big\rangle 
\Big]
  \nonumber  \\
 \hspace*{-20mm}
&& 
\hspace*{-18mm}
 +(1\!+\!\tilde{u}^{-2})\int\!{\rm D}z {\rm D}y_0\!\int\!\rmd t~ p(t |S\bra a\ket^{\frac{1}{2}}y_0, \lam^0)
W\Big(\tilde{u}^2\rme^{\tilde{u}^2+wy_0+vz}\Lambda(t)\Big)
\label{eq:E_gamma_infty}
\end{eqnarray}
What remains in our RS analysis  is to determine the order parameters $\{\tilde{u},v,w,\tilde{f},\tilde{g},\lambda\}$ by extremization of 
(\ref{eq:Psi_gamma_infty}), and to substitute the result into (\ref{eq:E_gamma_infty}). 
  
  \subsection{Scalar saddle point equations}
  
Partial   differentiation of (\ref{eq:Psi_gamma_infty}) with respect to the five scalar order parameters $\{\tilde{u},v,w,\tilde{f},\tilde{g}\}$ is now straightforward and gives, upon using identities such as $W^\prime(z)=W(z)/z[1+W(z)]$ and after manipulations similar to those used in  \cite{coolen2017replica}:
  \begin{eqnarray}
  \hspace*{-15mm}
  \zeta\tilde{f}\tilde{u}^4&=& -
\int\!{\rm D}z {\rm D}y_0\!\int\!\rmd t~ p(t |S\bra a\ket^{\frac{1}{2}}y_0, \lam^0)
\Big[
W\Big(\tilde{u}^2\rme^{\tilde{u}^2+wy_0+vz}\Lambda(t)\Big)\!-\!\tilde{u}^2\Big]^2
\label{eq:scalar_1}
\\
  \hspace*{-15mm}
   \zeta
 \tilde{g}\tilde{u}^2&=&  
\int\!{\rm D}z {\rm D}y_0\!\int\!\rmd t~ p(t |S\bra a\ket^{\frac{1}{2}}y_0, \lam^0) 
\frac{W\big(\tilde{u}^2\rme^{\tilde{u}^2+wy_0+vz}\Lambda(t)\big)}{1\!+\!W\big(\tilde{u}^2\rme^{\tilde{u}^2+wy_0+vz}\Lambda(t)\big)}
\label{eq:scalar_2}
\\
  \hspace*{-15mm}
  0&=& 
   \zeta w\Big[
 \bra a\ket  \Big\langle 
\frac{a^2}{2 \eta\! +\! \tilde{g}a}\Big\rangle^{\!-1}\!
-\tilde{g}
\Big]
\nonumber
\\
  \hspace*{-15mm}
&&
+\frac{1}{\tilde{u}^2}\int\!{\rm D}z {\rm D}y_0~y_0\!\int\!\rmd t~ p(t |S\bra a\ket^{\frac{1}{2}}y_0, \lam^0)
W\Big(\tilde{u}^2\rme^{\tilde{u}^2+wy_0+vz}\Lambda(t)\Big)
\label{eq:scalar_3}
\\
  \hspace*{-15mm}
\tilde{u}^2&=&
\Big\langle\frac{a}{2 \eta\! +\! \tilde{g}a} \Big\rangle 
\label{eq:scalar_4}
\\
  \hspace*{-15mm}
v^2&=&    
 w^2\Big[\bra a\ket  \Big\langle 
\frac{a^2}{2 \eta\! +\! \tilde{g}a}\Big\rangle^{\!-2}
\Big\langle 
\frac{a^3}{(2 \eta\! +\! \tilde{g}a)^2}\Big\rangle-1\Big]
-\tilde{f}\Big\langle\frac{a^2}{(2 \eta\! +\! \tilde{g}a)^2} \Big\rangle 
\label{eq:scalar_5}
  \end{eqnarray}
Compared to the simpler scenario of \cite{coolen2017replica}, the present RS theory involves two additional order parameters, $\tilde{f}$ and $\tilde{g}$. As a simple test we can set $\eta=0$, i.e. remove the priors for association parameters. This reduces the last two of the above saddle point equations  to
  $\tilde{g}=1/\tilde{u}^2$ and $\tilde{f}=-v^2/\tilde{u}^4$, removes all dependencies of the theory on the spectrum $\varrho(a)$ of the covariate correlation matrix (other than via $\bra a\ket$), and
   simplifies the remaining three scalar order parameter equations correctly to those derived in \cite{coolen2017replica}.

\subsection{Functional saddle point equation\label{sec:va}}

The equation from which to solve the functional order parameter $\lambda(t)$ is derived by functional differentiation of (\ref{eq:Psi_gamma_infty}). Upon using the short-hand 
$p(t)=\int\!{\rm D}y_0\!\int\!\rmd t~ p(t |S\bra a\ket^{\frac{1}{2}}y_0, \lam^0)$ for the typical distribution of the event times in the data, this equation takes the form
\begin{eqnarray}
\hspace*{-15mm}
\frac{p(t)}{\lambda(t)}  &=&
\int\!{\rm D}z {\rm D}y_0\!\int_t^\infty\!\!\frac{\rmd t^\prime}{\tilde{u}^2\Lambda(t^\prime)}~ p(t^\prime |S\bra a\ket^{\frac{1}{2}}y_0, \lam^0)
W\Big(\tilde{u}^2\rme^{\tilde{u}^2+wy_0+vz}\Lambda(t^\prime)\Big)
\end{eqnarray}
It differs only minimally from the one in  \cite{coolen2017replica}, and is hence equally difficult to solve analytically. Following \cite{coolen2017replica} we will therefore follow a variational approach, motivated by the asymptotic form of the solution for large times (see \cite{coolen2017replica}  for details), and choose the functional ansatz $\Lambda(t) = k [\Lambda^0(t)]^{\rho}$, leaving two variational parameters $(k,\rho)$ to be solved, instead of a function. Inserting this  ansatz into (\ref{eq:Psi_gamma_infty}), followed by partial differentiation with respect to $k$ and $\rho$ then leads to the following two equations:
\begin{eqnarray}
\hspace*{-20mm}
 \tilde{u}^2&=&
\int\!{\rm D}z {\rm D}y_0\!\int\!\rmd t~ p(t |S\bra a\ket^{\frac{1}{2}}y_0, \lam^0)
W\Big(k\tilde{u}^2\rme^{\tilde{u}^2+wy_0+vz} [\Lambda^0(t)]^{\rho}\Big)
\\
\hspace*{-20mm}
 0 &=&
\frac{1}{\tilde{u}^2}\int\!{\rm D}z {\rm D}y_0\!\int\!\rmd t~ p(t |S\bra a\ket^{\frac{1}{2}}y_0, \lam^0)
W\Big(k\tilde{u}^2\rme^{\tilde{u}^2+wy_0+vz} [\Lambda^0(t)]^{\rho}\Big)\log \Lambda^0(t)
  \nonumber  \\
 \hspace*{-20mm}
&&
-\frac{1}{\rho}
-\int\!{\rm D}y_0\!\int\!\rmd t~ p(t |S\bra a\ket^{\frac{1}{2}}y_0, \lam^0)
\log\Lambda^0(t)
\end{eqnarray}
These have to be solved numerically alongside (\ref{eq:scalar_1}-\ref{eq:scalar_5}). We will compactify our equations by using instead of $k$ the variable $q=k\tilde{u}^2 \exp(\tilde{u}^2)$. As a further benefit of our variational ansatz, the time integrations in the saddle point equations can be simplified significantly upon switching to the new integration variable $s=\exp[-\exp(S\bra a\ket^{\frac{1}{2}}y_0)\Lambda^0(t)]\in[0,1]$, which gives $\rmd s=- p(t |S\bra a\ket^{\frac{1}{2}}y_0, \lam^0)\rmd t$. After this transformation we can also combine the Gaussian variables into a single one, giving
 \begin{eqnarray}
  \hspace*{-20mm}
  \zeta\tilde{f}\tilde{u}^4&=& -
\int\!{\rm D}x \int_0^1\!\rmd s~ 
\Big[
W\big(q\rme^{\sigma x}\log^\rho(1/s)\big)\!-\!\tilde{u}^2\Big]^2
\label{eq:scalar_1_final}
\\
  \hspace*{-20mm}
   \zeta
 \tilde{g}\tilde{u}^2&=&  
\int\!{\rm D}x \int_0^1\!\rmd s~ 
\frac{W\big(q\rme^{\sigma x}\log^\rho(1/s)\big)}{1\!+\!W\big(q\rme^{\sigma x}\log^\rho(1/s)\big)}
\label{eq:scalar_2_final}
\\
  \hspace*{-20mm}
w&=&  \frac{\tilde{g}\rho S}{ \bra a\ket^{\frac{1}{2} }}\Big\langle 
\frac{a^2}{2 \eta\! +\! \tilde{g}a}\Big\rangle
\label{eq:scalar_3_final}
\\
  \hspace*{-20mm}
\tilde{u}^2&=&
\Big\langle\frac{a}{2 \eta\! +\! \tilde{g}a} \Big\rangle 
\label{eq:scalar_4_final}
\\
  \hspace*{-20mm}
v^2&=&    
 w^2\Big[\bra a\ket  \Big\langle 
\frac{a^2}{2 \eta\! +\! \tilde{g}a}\Big\rangle^{\!-2}
\Big\langle 
\frac{a^3}{(2 \eta\! +\! \tilde{g}a)^2}\Big\rangle-1\Big]
-\tilde{f}\Big\langle\frac{a^2}{(2 \eta\! +\! \tilde{g}a)^2} \Big\rangle 
\label{eq:scalar_5_final}
\\
\hspace*{-20mm}
 \tilde{u}^2&=&
\int\!{\rm D}x\!\int_0^1\!\rmd s~ W\big(q\rme^{\sigma x}\log^\rho(1/s)\big)
\label{eq:scalar_6_final}
\\
\hspace*{-20mm}
 \frac{\tilde{u}^2}{\rho}&=&
\int\!{\rm D}x \!\int_0^1\!\!\rmd s~ 
W\Big(q\rme^{\sigma x}\log^\rho(\frac{1}{s})\Big)
\log\log(\frac{1}{s})
- \zeta \tilde{g}\tilde{u}^2S\bra a\ket^\frac{1}{2}(w\!-\!\rho S\bra a\ket^\frac{1}{2}) 
+\tilde{u}^2 C_{\rm E}
\nonumber
\\
\hspace*{-20mm}
\label{eq:scalar_7_final}
  \end{eqnarray}
in which $\sigma^2=(w-\rho S\bra a\ket^{\frac{1}{2}})^2+v^2$, $C_{\rm E}$ denotes Euler's constant, and where we used the integral $\int_0^1\!\rmd s~\log\log(1/s)=\int_0^\infty\!\rmd x~\rme^{-x}\log x=-C_{\rm E}$.
%%%%%%%%%%%%%%%%%%%%%%%%%%%%%%%%%%

\subsection{The limits $\eta\to 0$, $\zeta\to 0$ and $\zeta\to\infty$}

Here we investigate the order parameter behaviour in the small and large $\zeta$ limits, to confirm the shape of the order parameter plots, both analytically  and by numerical analysis in the next section. In particular, since the order parameters $v$ and $w$ increase with $\zeta$ for small $\zeta$ and tend to zero for large $\zeta$, we conclude that there must be a stationary point between these two extremes. This analytical argument is validated by numerical solutions of \eref{eq:scalar_1_final}-\eref{eq:scalar_7_final} (see \fref{fig:op_wve}) and by synthetic data studies (\fref{fig:predicted_slope}).  An explanation of this phenomenon in terms of model complexity and the emergence of statistical constraints can be constructed \cite{bulso2019complexity}.
In the limit $\eta\to 0$, describing a fully flat prior for association parameters, the regression changes from MAP to ML, and our RS equations should therefore reduce to those of  \cite{coolen2017replica}. Upon setting $\eta\to 0$ in 
equations  (\ref{eq:scalar_1_final}--\ref{eq:scalar_7_final}), we immediately find that 
\begin{eqnarray}
w=  \rho S \bra a\ket^{\frac{1}{2}},~~~~~~
\tilde{g}=1/\tilde{u}^2,~~~~~~
\tilde{f}=-v^2/\tilde{u}^4
\end{eqnarray}
From the first of these it follows that $\sigma=v$, and that the remaining RS scalar order parameter equations from which to solve $\{v,\rho,\tilde{u},q\}$ hence simplify to
 \begin{eqnarray}
  \zeta v^2&=& 
\int\!{\rm D}x \int_0^1\!\rmd s~ 
\Big[
W\big(q\rme^{v x}\log^\rho(1/s)\big)\!-\!\tilde{u}^2\Big]^2
\\
  \hspace*{-0mm}
   \zeta
&=&  
\int\!{\rm D}x \int_0^1\!\rmd s~ 
\frac{W\big(q\rme^{v x}\log^\rho(1/s)\big)}{1\!+\!W\big(q\rme^{v x}\log^\rho(1/s)\big)}
\\
\hspace*{-0mm}
 \tilde{u}^2&=&
\int\!{\rm D}x\!\int_0^1\!\rmd s~ W\big(q\rme^{v x}\log^\rho(1/s)\big)
\\
\hspace*{-0mm}
 \frac{\tilde{u}^2 }{\rho}&=&
\int\!{\rm D}x \!\int_0^1\!\!\rmd s~ 
W\Big(q\rme^{v x}\log^\rho(\frac{1}{s})\Big)
\log\log(\frac{1}{s})
+\tilde{u}^2 C_{\rm E}
  \end{eqnarray}
  We observe that, as a consequence of having modified our present derivation compared to the one in \cite{coolen2017replica} (we changed the order of integral transformations and partial differentiations), the above expressions provide the proof for the simplifying identity $w=  \rho S$ (which holds for the case where $\bra a\ket=1$), that was suggested by numerical analysis but not yet proven in  \cite{coolen2017replica}. For $\eta\to 0$ we can thus retrieve from our present results in an even more satisfactory manner the variational RS theory of  \cite{coolen2017replica}. 

For $\zeta\to 0$ (no overfitting) we expect to find $v\to 0$ and $w,\rho,k\to 1$. In analogy with 
 \cite{coolen2017replica} we now make the ans\"{a}tze that $\tilde{u},v=O(\sqrt{\zeta})$ and $\rho=1+O(\zeta)$ for $\zeta\to 0$, and expand our equations (\ref{eq:scalar_1_final}--\ref{eq:scalar_7_final}) in leading order for small $\zeta$, using $W(z)=z+O(z^2)$ for $z\to 0$. After expanding the various integrals, whose leading orders in $\zeta$ can all be done analytically, this results in
 \begin{eqnarray}
 \hspace*{-10mm}&&
  \tilde{u}^2/\zeta=1+O(\zeta),~~~~~~
   \zeta
 \tilde{g}=  1
+O(\zeta),~~~~~~
w=  S \bra a\ket^{\frac{1}{2}}+O(\zeta),\\
 \hspace*{-10mm}
&& \tilde{f}\zeta=-1+O(\zeta),~~~~~~~k=1+O(\zeta),~~~~~~~
 v^2/\zeta= 
    1
    +O(\zeta),
  \end{eqnarray}
  which confirms that in the absence of overfitting we indeed recover the correct values of the order parameters from our RS equations.

  Finally we inspect the behaviour of the RS equations (\ref{eq:scalar_1_final}--\ref{eq:scalar_7_final}) in the limit $\zeta\to\infty$ of a diverging imbalance between the number of covariates and the number of samples. Note that in  \cite{coolen2017replica} (i.e. for $\eta=0$) 
  this limit was inaccessible, due to a phase transition at $\zeta=1$, where $v,w\to\infty$. In the present theory, describing the regularized version of the Cox model,  this phase transition is suppressed by the Bayesian prior, provided we choose $\eta>0$. We now make the ansatz that $\tilde{g}\to 0$ for $\zeta\to\infty$, giving $\tilde{u}^2\to \bra a\ket/2\eta$, $v\to 0$, $w\to 0$, $\tilde{f}\to 0$, $\sigma^2\to \rho^2 S^2\bra a\ket$,  and upon introducing $Q=\lim_{\zeta\to\infty}   \zeta
 \tilde{g}\tilde{u}^2$, the remaining trio $\{Q,q,\rho\}$ is for $\zeta\to\infty$ to be solved from the remaining three coupled equations
   \begin{eqnarray}
   \hspace*{-20mm}
~~Q&=&  
\int\!{\rm D}x \int_0^1\!\rmd s~ 
\frac{W\big(q\rme^{\rho S
\bra a\ket^{
\frac{1}{2}} x}\log^\rho(1/s)\big)}{1\!+\!W\big(q\rme^{\sigma x}\log^\rho(1/s)\big)}
\\
\hspace*{-20mm}
 \frac{\bra a\ket}{2\eta}
 &=&
\int\!{\rm D}x\!\int_0^1\!\rmd s~ W\big(q\rme^{\rho S
\bra a\ket^{
\frac{1}{2}} x}\log^\rho(1/s)\big)
\\
\hspace*{-20mm}
 \frac{\bra a\ket}{2\eta\rho}&=&
\int\!{\rm D}x \!\int_0^1\!\!\rmd s~ 
W\Big(q\rme^{\rho S
\bra a\ket^{
\frac{1}{2}} x}\log^\rho(\frac{1}{s})\Big)
\log\log(\frac{1}{s})
+ Q S^2\bra a\ket\rho 
+\frac{\bra a\ket}{2\eta}C_{\rm E}
  \end{eqnarray}
 For $\zeta\to\infty$ we thus expect to  find, as a consequence of  $\lim_{\zeta\to \infty}v=\lim_{\zeta\to\infty}w= 0$, vanishing inferred association parameters in the present regularized Cox model, with the assumed scaling of the width of the prior.

\subsection{Expression for the overfitting measure }

Finally, 
using the variational approximation for the cumulative hazard rate, the simple manipulations applied to the RS saddle point equations,  and the actual order parameter equations themselves, the overfitting measure (\ref{eq:E_gamma_infty}) can be simplified to the transparent form
\begin{eqnarray}
\hspace*{-10mm}
\lim\limits_{N \to \infty} E_{\infty}(\bb^0, \lam^0) &=&
\eta\zeta\Big[
w^2\bra a\ket  
 \Big\langle \frac{a^2}{2 \eta\! +\! \tilde{g}a}\Big\rangle^{\!-2}
 \Big\langle  \frac{a^2}{(2 \eta \! +\! \tilde{g}a)^2}\Big\rangle
- \tilde{f}\Big\langle\frac{a}{(2 \eta\! +\! \tilde{g}a)^2} \Big\rangle 
\Big]
\nonumber
\\
\hspace*{-10mm}
&&
\hspace*{10mm}
+\int\!\rmd t~ p(t)
\log\Big( \frac{\lambda^0(t)}{\lambda(t)}\Big)-\zeta\eta  S^2  
\end{eqnarray}
with the short-hand $p(t)=\int\!{\rm D}y_0~p(t|S\bra a\ket^{\frac{1}{2}}y_0,\lambda^0)$. 
Our  variational ansatz $\Lambda(t)=k[\Lambda^0(t)]^\rho$ implies that 
$\lambda(t)=
k\rho\lambda^0(t)[\Lambda^0(t)]^{\rho-1}$, 
hence
\begin{eqnarray}
\hspace*{-15mm} 
\int\!\rmd t~ p(t)
\log\Big( \frac{\lambda^0(t)}{\lambda(t)}\Big)&=& 
-\log k -\log\rho
-(\rho\!-\!1)\int\!\rmd t~ p(t)\log \Lambda^0(t)
\nonumber
\\
\hspace*{-15mm} 
&=& -\log k -\log\rho
-(\rho\!-\!1) \int\!{\rm D}y_0\int_0^1\!\rmd s\Big[\rme^{-S\bra a\ket^{\frac{1}{2}}y_0}\log(\frac{1}{s})\Big]
\nonumber
\\
\hspace*{-15mm} 
&=& -\log k -\log\rho
-(\rho\!-\!1)\int_0^\infty\!\rmd x~\rme^{-x}\log x
\nonumber
\\
\hspace*{-15mm} 
&=& -\log k -\log\rho
+(\rho\!-\!1)C_{\rm E}
\end{eqnarray}
Our final result for the asymptotic overfitting measure $E(S)=\lim_{N\to\infty}E_{\infty}(\bb^0, \lam^0) $ is therefore
\begin{eqnarray}
E(S) &=& 
\eta\zeta\Big[
w^2\bra a\ket  
 \Big\langle \frac{a^2}{2 \eta\! +\! \tilde{g}a}\Big\rangle^{\!-2}
 \Big\langle  \frac{a^2}{(2 \eta \! +\! \tilde{g}a)^2}\Big\rangle
- \tilde{f}\Big\langle\frac{a}{(2 \eta\! +\! \tilde{g}a)^2} \Big\rangle 
\Big]
\nonumber
\\
&&
\hspace*{10mm}
-\log k -\log\rho
+(\rho\!-\!1)C_{\rm E}-\zeta\eta  S^2  
\end{eqnarray}
We observe that, as was the case in  \cite{coolen2017replica} (without regularization), both the RS order parameter equations and the overfitting measure have within the variational approximation become completely independent of the true base hazard rate $\lambda^0(t)$.

\section{Numerical experiments\label{sec:numerical}}

\subsection{Numerical solution of order parameter equations}

\begin{figure}[t]
\hspace*{-5.5mm}
\includegraphics[width=7.5cm]{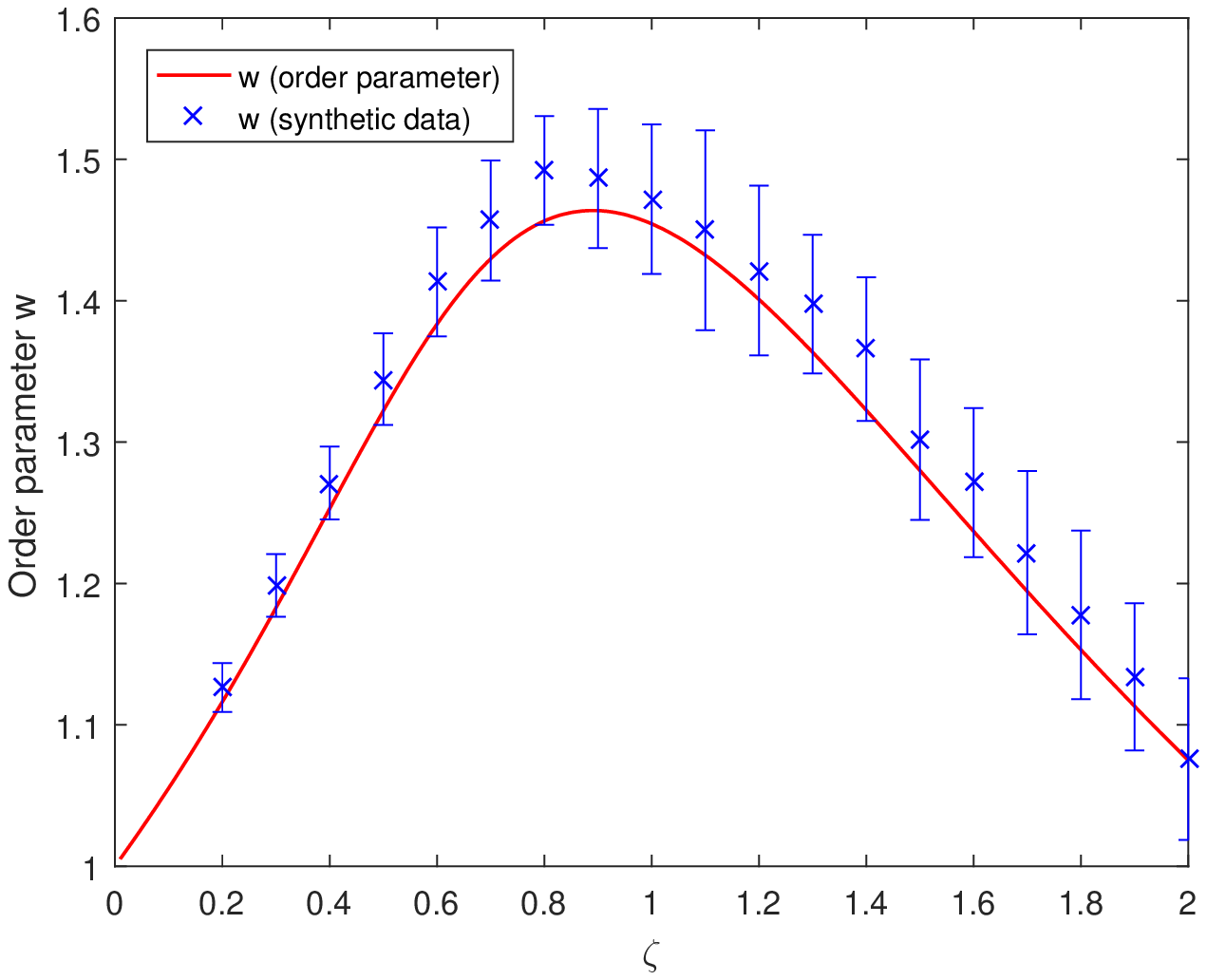}\hspace*{-5mm}\includegraphics[width=7.5cm]{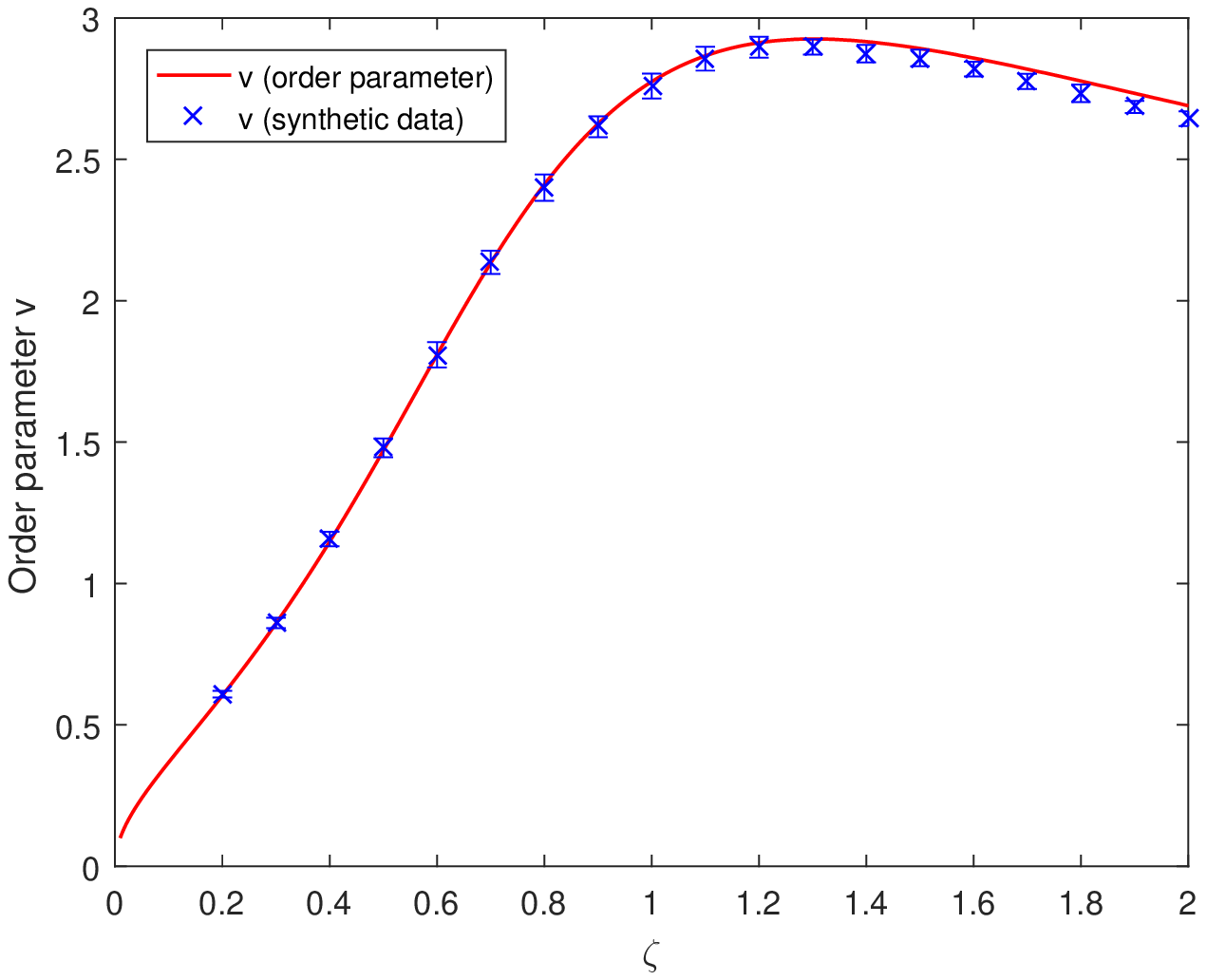} 
\caption{
Predicted and measured values  of the order parameters $w$ and $v$ (solid lines and markers, respectively), 
for $\A = \I$, $S=1$ and  $p=2000$, shown versus $\zz=p/N \in (0,2]$. Measurements are determined via MAP regression, with regularization parameter $\eta = 0.025$. Simulations are repeated 50 times with independent data sets (generated according to \cite{bender2005generating}, with constant hazard rates), and results shown as averages with error bars indicating one standard deviation. Note that for these settings, slope and the width of the association parameter cloud equal $w$ and $v$, respectively. 
 }
\label{fig:op_wve}
\end{figure}

\begin{figure}[t]
\hspace*{-5.5mm}
\includegraphics[width=7.5cm]{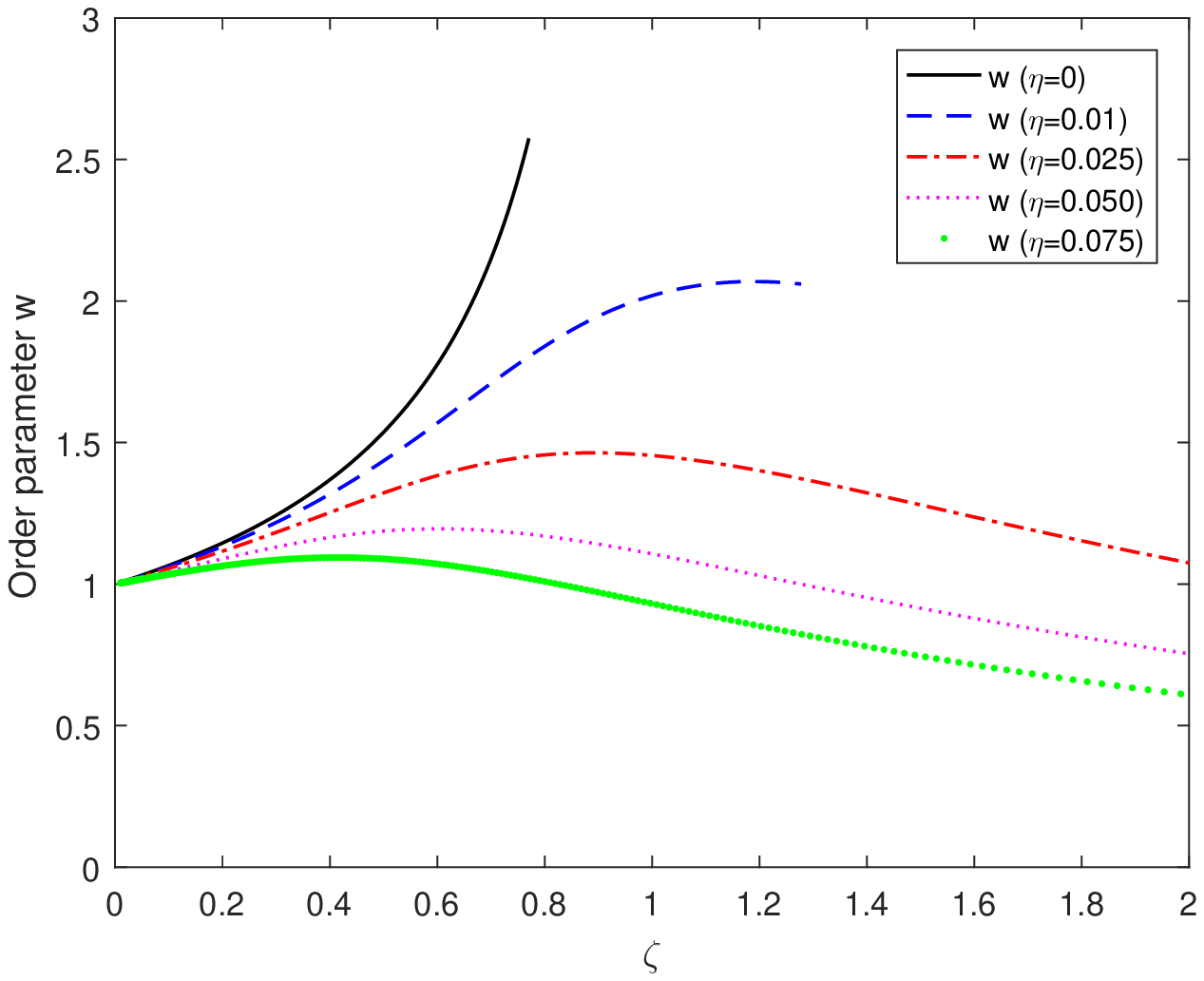}\hspace*{-7mm}\includegraphics[width=7.5cm]{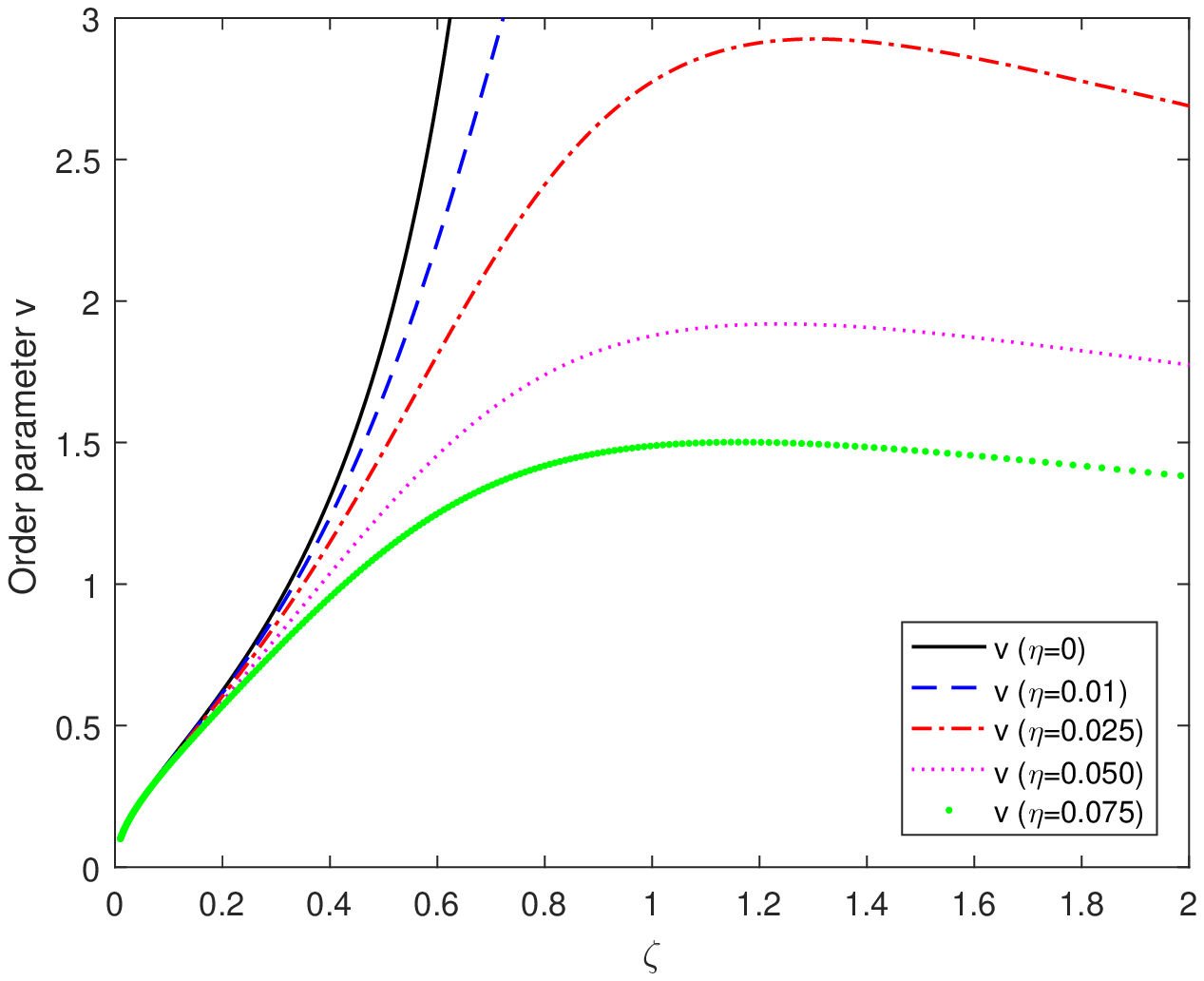}  \\
\vspace*{-3mm}
\caption{
Predicted values of the order parameters $w$ (left) and $v$ (right), shown versus $\zz=p/N$. They are  obtained by solving numerically the RS equations (\ref{eq:scalar_1_final}--\ref{eq:scalar_7_final})  
for $\A = \I$ and $S=1$, with the variational approximation for $\lambda(t)$, and different choices of the regularization parameter $\eta$. 
}
\label{fig:predicted_slope}
\end{figure}

Numerical solution of the RS saddle point equations  (\ref{eq:scalar_1_final}--\ref{eq:scalar_7_final}), with the variational approximation for the base hazard rate, results in data as shown in Figure \ref{fig:op_wve}. This Figure corresponds to  $\A = \I$, i.e. uncorrelated and normalized covariates, and $S=1$.  The phase transition at $\zeta=1$ of \cite{coolen2017replica} (corresponding to ML regression) is for $\eta>0$ no longer present, due to the regularization incorporated into MAP regression. As $\eta$ increases, we find the slope $\kappa$ (which for the present parameter settings is identical to $w$) and the variance $v$ of the data cloud decreasing.  The increase of $E$ with $\eta$ indicates that also overfitting  is reduced by regularization. 

To test the above predictions, 
we generated synthetic time-to-event data using zero mean covariate vectors $\z$ with covariance matrix $\A$, and Gaussian random and zero-average association vectors $\bb^0$,  for different values of $N$ and $p$. Base hazard rates were chosen to be constant. Event times were generated from the Cox proportional hazards model following \cite{bender2005generating}, and from the simulated data we then extracted estimates of the association parameters via penalized Cox regression (using the R package, \emph{glmnet} \cite{friedman2009glmnet}). Upon solving our RS order parameter equations (\ref{eq:scalar_1_final}--\ref{eq:scalar_7_final}) for the chosen values of $\zeta=p/N$ and $S^2=p^{-1}(\bb^0)^2$, we compared the solution with the regression outcomes  via (\ref{eq:physical_meaning}), under various conditions.
By construction, there is no model mismatch, since the data are generated from the model assumed in parameter inference. 
Our theoretical predictions for the slope and variance agree remarkably well with the simulations; see figure \ref{fig:predicted_slope}. 

\begin{figure}[t]
\hspace*{-5.5mm}
\includegraphics[width=7.5cm]{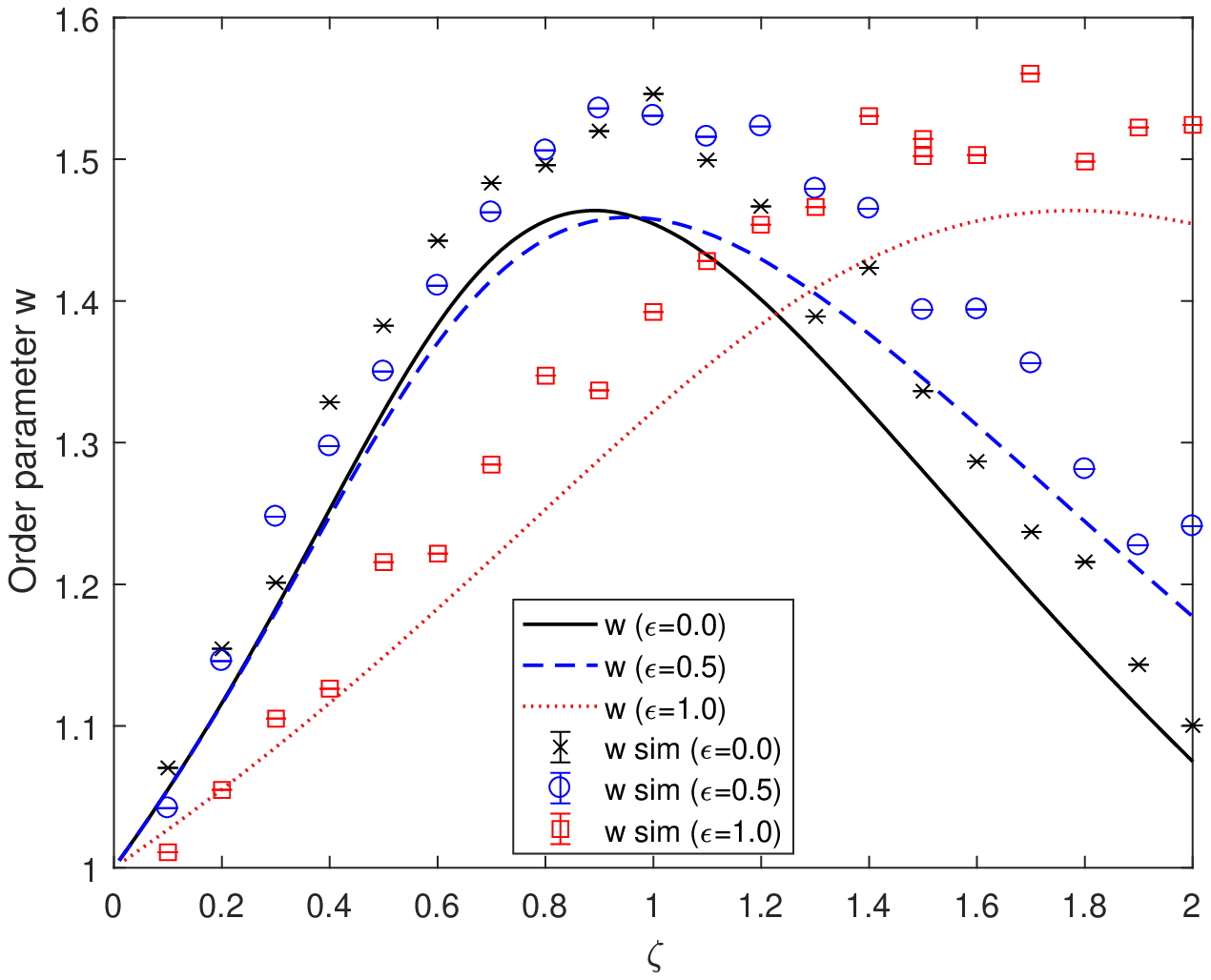}\hspace*{-6mm}\includegraphics[width=7.5cm]{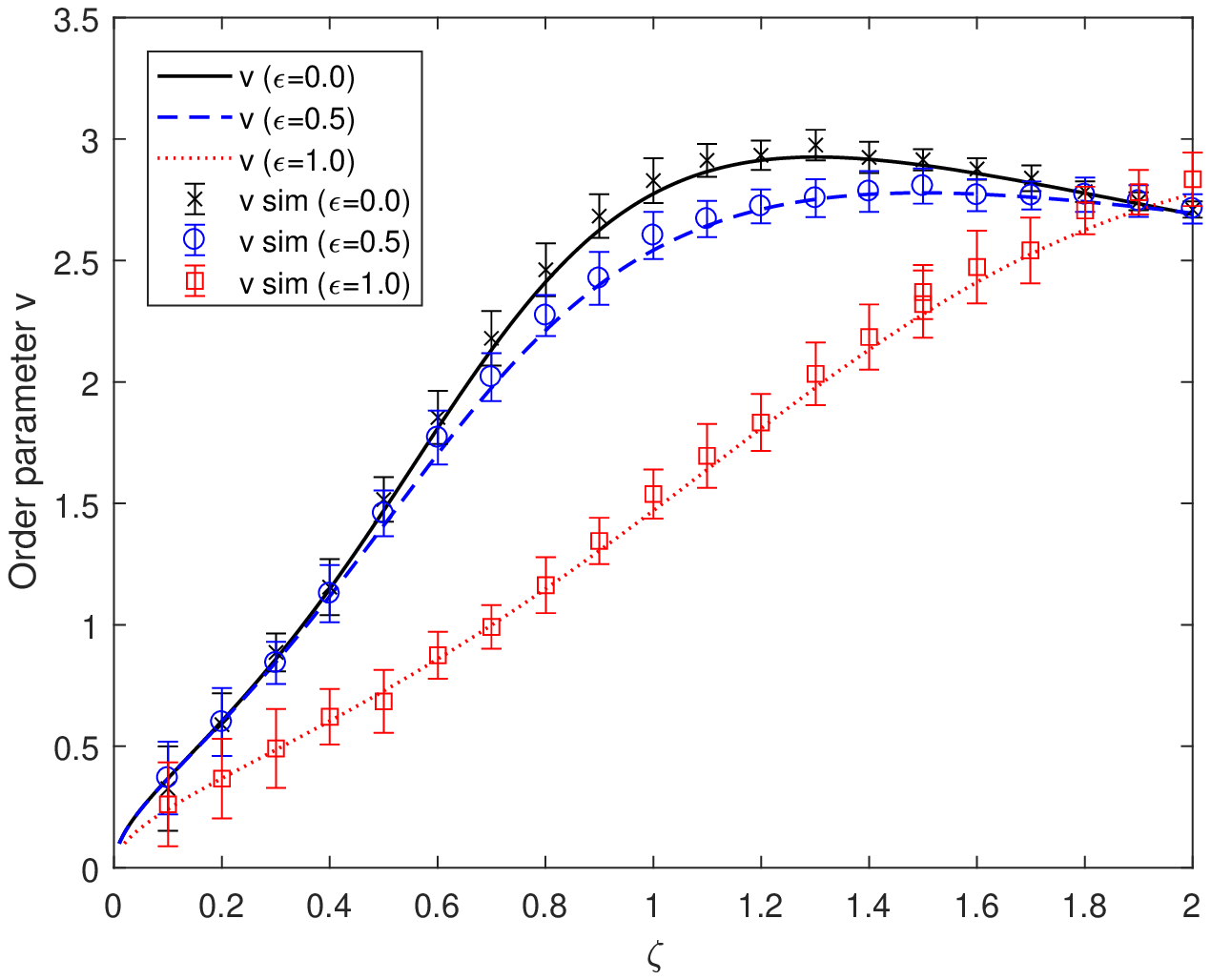}
\caption{Predicted values of the order parameters $w$ (left) and $v$ (right), shown versus $\zz=p/N$. They are obtained by solving numerically the RS equations (\ref{eq:scalar_1_final}--\ref{eq:scalar_7_final})  
for $\eta=0.025$ and $S=1$, with the variational approximation for $\lambda(t)$. Here the covariates are pairwise correlated according to  
$A_{\mu, \mu+1}=A_{\mu+1,\mu}=\epsilon$ for all $\mu$ odd, with $A_{\mu\nu}=0$ for all other $\mu\neq \nu$, with $\epsilon\in[0,1]$. Note that for these settings, $w$ and $v$ are the slope and the width of the association data cloud. For the left $w$ plot, only mean simulation values are shown since including the errors bars of approximately $\pm 10\%$ led to cluttered plots. Error bars can be displayed clearly for all values of $\epsilon$ on the right $v$ plot. The markers each represent averages over 32 regressions with distinct covariate and association realizations and we fix the value of $Np=400,000$. 
} 
\label{fig:w_correl}
\end{figure}

The effect of covariate collinearity on the inferred regression coefficients \cite{stewart1987collinearity} was investigated with two non-diagonal  covariance matrices $\A$,  both with $\lim_{p\to\infty}\bra a\ket=1$ and $\lim_{p\to\infty}\bra a^2\ket=1+\epsilon^2$ (hence with spectra of finite width), and $\epsilon={\mathcal O}(1)$.  This ensures that the requirements for self-averaging of the RS theory on the eigenvalue spectrum $\varrho(a)$ of $\A$ are fulfilled.  Our first choice was $A_{\mu\nu} = \delta_{\mu\nu} + (1\!-\! \delta_{\mu\nu})\epsilon/\sqrt{p}$, with   eigenvalues 
$1\!-\!\epsilon/\sqrt{p}$ (multiplicity $p\!-\!1$) and $1\!+\!(p\!-\!1)\epsilon/\sqrt{p}$ (multiplicity $1$).  Upon working out the spectrum-dependent quantities in the RS equations, we find that for this matrix choice they are independent of $\epsilon$. Hence the order parameters are predicted to be identical  to those for data with uncorrelated covariates. Simulations (not shown here) confirm that this is indeed the case, modulo finite size fluctuations. Our second choice for $\A$ had again $A_{\mu\mu}=1$ for all $\mu$, but now covariates are correlated in ordered pairs: $A_{\mu, \mu+1}=A_{\mu+1,\mu}=\epsilon$ for all $\mu$ odd, with $A_{\mu\nu}=0$ for all other $\mu\neq \nu$ (with $ 0\leq \epsilon\leq 1$). This is a block diagonal matrix with $\varrho(a)=\frac{1}{2}\delta(a\!-\!1\!-\!\epsilon)+\frac{1}{2}\delta(a\!-\!1\!+\!\epsilon)$, and the RS order parameters {\em will} depend on the strength $\epsilon$ of the covariate correlations.
In \fref{fig:w_correl}, we show the values of the order parameters $v$ and $w$, as solved from the RS equations, for $S=1$, $\eta=0.025$ and different values of the correlation parameter $\epsilon$, as functions of $\zeta$. Here we again have $\kappa=w$.
In the same figure we show the results of numerical simulations carried out for $\epsilon = \{0.0, 0.5, 1.0\}$ and $Np=400,000$. The error bars of approximately $\pm 10\%$ were not displayed for clarity. The covariates were generated according to: $z^{i}_{\mu}=y_{i\mu}$ for $\mu$ odd, and $z^i_{\mu}=\epsilon y_{i \mu-1}+\sqrt{1\!-\!\epsilon^2}y_{i\mu}$, in which all $\{y_{i\mu}\}$ are independent Gaussian random variables, with $\bra y_{i\mu}\ket=0$ and $\bra y_{i\mu}^2\ket=1$.
This choice generates the above covariate correlations $A_{\mu\nu}$. 
The markers each represent averages over 32 regressions with distinct covariate and association realizations. The agreement between theory and simulations is seen to be quite satisfactory. We observe that the effect of covariate correlations on the overfitting noise is always a reduction ($v$ decreases with $\epsilon$).
%Standard deviations are not shown, to prevent cluttering of the figure, but are of the order of $\Delta v\approx 0.05$ and $\Delta w\approx 0.1$. 
% For $\zeta<1$ we observe that covariate correlations reduce the inflation of association parameters caused by overfitting ($w$ decreases slightly with $\epsilon$), whereas for $\zeta>1$ they increase this inflation ($w$ increases). Their effect on the overfitting noise is always a reduction ($v$ decreases with $\epsilon$).

Covariates of real survival data can obviously be distributed in many different ways. The assumption in \eref{eq:y_mvn} of Gaussian distributed risk scores is a direct consequence of working in the  limit $p\to\infty$, in combination with  the Central Limit Theorem. More specifically, there is no need to assume Gaussian covariate statistics. To verify the validity of Gaussian risk score statistics, we carried out simulations with four common covariate distributions, all with identical first two moments: normal, $p(z_i) = \mathcal{N}(0,1)$, Rademacher, $p(z_i) = \half \delta(z_i-1) + \half \delta(z_i+1)$, uniform, $z_i \sim \mathcal{U}(-\sqrt{3}, \sqrt{3})$ and the student t-distribution, $z_i \sim{\rm t\!-\!dist}(\nu) / \sqrt{\nu/(\nu\!-\!2)}$ (with degrees of freedom $\nu=5$). The deviations between predictions using Gaussian covariates and the above distributions were indeed small ($<1\%$ for $w$ and $<0.5\%$ for $v$) validating our asymptotic assumption that our theory admits a range of covariates distributions.
\vsp

\begin{figure}[t]
\begin{center}
\includegraphics[width=9.5cm,angle=0]{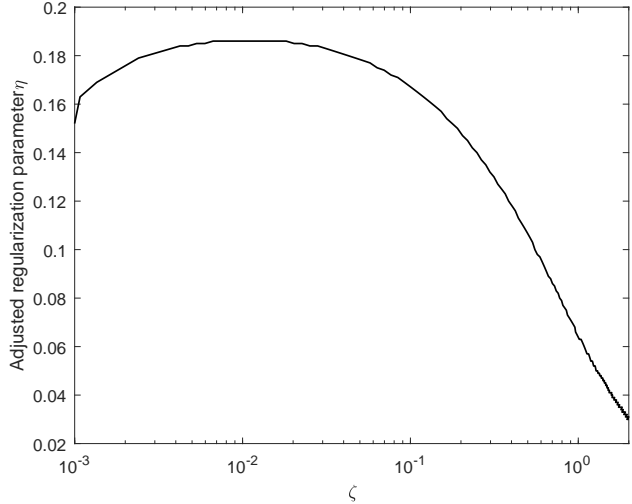}
 \end{center}\vspace*{-2mm}
\caption{The present theory allows for the analytical identification of the optimally adjusted MAP regularization parameter for Cox regression, by solving the RS order parameter equations (\ref{eq:scalar_1_final}--\ref{eq:scalar_7_final})  upon demanding unbiased recovery of regression coefficients, $\kappa=1$, with $\eta$ as parameter to be solved instead of $w$. Here we show the result versus $\zeta=p/N$. It is not straightforward to solve the order parameter equations close to $\zz=0$,  but we know that  the curve should tend to the origin for $\zeta=0$ (where ML inference is asymptotically exact). }
\label{fig:auto}
\end{figure}

In MAP analyses the regularization parameter $\eta$ is usually determined by $k$-fold cross-validation, or via the Generalized Cross Validation (GCV) estimator \cite{craven1978smoothing}. A fraction of the data is set aside for this purpose, leaving fewer samples available for inference of model parameters. This has a detrimental effect on inference accuracy. Our present theory, in contrast, suggests a more data efficient method of estimating the amount of regularization needed, without the need to sacrifice any samples. 
By fixing the slope parameter to unbiased recovery of the regression coefficients, i.e. $w/\tS=1$, and  solving the order parameter equations (\ref{eq:scalar_1_final}--\ref{eq:scalar_7_final}) with $\eta$ as a parameter to be determined (instead of $w$), the {\em optimal} values of $\eta$ can be estimated without any cross-validation; see Figure \ref{fig:auto}. The optimal values in Figure \ref{fig:auto} are seen to match those $(\zeta, \eta)$ pairs in Figure \ref{fig:op_wve} where $w/\tS=1$, as they should. For example, when $\zz=1$, the required amount of regularization to compensate for high covariate dimensionality  can be read off from Figure \ref{fig:auto} to be  $\eta\approx 0.05$. 
In interpreting this figure, however, we should note our rescaling of our association parameters, prompted by the observation that $\bb^2=O(1)$ is required to avoid non-finite event times for large $p$. This implied that our $L_2$ prior in MAP inference is of the form $p(\bb)\propto \exp(-\eta p\bb^2)$. 

The upward sloping region of Figure \ref{fig:auto}, for small $\zeta$, matches our intuition of requiring an increasing amount of regularization for an increasing $\zz$ (up to $\zz \approx 0.01$).  However, as $\zz$ is increased further, we see that optimal regularization now requires a decreasing value of $\eta$.  To test this less intuitive prediction, we chose four larger values of $\zz$, read off the required values of $\eta$ for unbiased inference from Figure \ref{fig:auto}, and calculated the slope of the association parameter cloud  from $100$ simulations. These predictions were made with $p=250$ suggesting our theory is valid for relatively low values of $p$ and $N$. The results show that the slope of the association parameter cloud is indeed unity, i.e. for the $\eta$ values proposed by the RS theory, the overfitting-induced inference bias is indeed suppressed as predicted; see the table below:
\\[0mm]
\begin{center}
 \begin{tabular}{|c | c | c | c|} 
 \hline
 $\zz$ & required $\eta$ & corresponding \emph{glmnet} $\lambda$ & mean slope $\pm$ 1 s.d \\ [0.5ex] 
 \hline \hline
0.110 &    0.165  &  0.036 & 1.007 $\pm$ 0.028 \\ 
 \hline
0.552 & 0.100 &  0.110 & 1.009 $\pm$ 0.081\\
 \hline
 1.055 & 0.062 &  0.131 & 1.013 $\pm$ 0.094\\
 \hline
 2.001 & 0.031 &  0.124 & 0.956 $\pm$ 0.139\\
 \hline
\end{tabular}
\end{center}
\vspace*{4mm}

\noindent
Note that Figure \ref{fig:auto}, together with our confirmation in regression simulations that the predicted optimal values of $\eta$  indeed induce unbiased MAP estimators for regression coefficients (i.e. slopes $\kappa=1$ in the association parameter clouds), confirm a posteriori the correctness of the chosen scaling with $p$ of our $L_2$ prior $p(\bb)\propto\exp(- \eta p\bb^2)$.
In those situations where the conditions for our theory to apply are not met, other properties may of course affect the optimal value of $\eta$. For instance, our simulated data are generated from the Cox model where the ground truth association vector $\bb^\star$ is not sparse. Equally, the data could be generated from a model with fewer nonzero associations, including choices for which the Central Limit Theorem no longer guarantees that the risk scores $\bb^\star\cdot\z$ have Gaussian statistics. 

%%%%%%%%%%%%%%%%%%%%%%%%%%%%%%%%%%%%%%%%%%%%%%%%%%%%%%%%%%%%%%%%%%%%%%%%%%

\section{Discussion\label{sec:discussion}}

Failure to correct multivariate ML or MAP regression results for overfitting can lead to serious inference errors. 
The inferred regression coefficients of the multivariate Cox model are known to be increasingly biased as the ratio of data dimension $p$ to the sample size $N$ increases. For medical time-to-event analysis, where it is possible to obtain (and common to have) large numbers of measurements per patient, such as genomic, epigenetic and imaging covariates, this bias is quite problematic. It induces false positive associations, which will inevitably turn out to be non-reproducible. This leads to a preventable waste of time and health funds, and frustrates the translation of  the significant progress made in recent decades in medical data acquisition into effective data-driven personalized medicine. In this paper, which builds on the recent study  \cite{coolen2017replica}, we  have  built successfully a theory to predict this bias for the multivariate Cox model in the presence of ridge regularization, when the data dimension scales as $p \sim N$. This paves the way further for effective overfitting corrections in multivariate MAP inference.  Alternatively, our analysis allows for a straightforward analytical determination of the optimal regularization needed to correct the overfitting bias, without having to sacrifice valuable training data to cross-validation.
In addition to overfitting-induced inference bias, there is a further effect of overfitting on inferred error bars.  To determine the statistical significance of inferred regression coefficients, p-values are typically used. These rely on asymptotic results which do not hold in the regime where both $p$ and $N$ are large with $\zz=p/N \sim \mathcal{O}(1)$  \cite{sur2018modern}\cite{fan2017nonuniformity}, leading to incorrect rejections of the null hypothesis. Our theory shows that the variance of the inferred regression coefficients around the true value is a function of $p/N$, necessitating an adjustment to traditional test statistics used in p-value calculations.

The aim of our theory is to provide epidemiologists and clinical trials practitioners with a means of analysing data where $p \sim \mathcal{O}(N)$. This paper considers a student-teacher learning problem where we assume the data-generating model is known. A practical overfitting correction protocol for multivariate MAP regression on high-dimensional time-to-event data, based on our present theory, requires  knowledge of the values of $S$ (the magnitude $|\bb^0|$ of the true association parameter vector) and of the eigenvalue spectrum of the covariate correlation matrix $\A$. For synthetic data, these are available by assumption. For real data, $S$ can be computed from the inferred regression parameters $\hat{\bb}$, alongside the RS order parameters, using (\ref{eq:physical_meaning}), from which one infers the relation $v^2\!+\!w^2=\hat{\bb}\cdot\A\hat{\bb}$ (in non-rescaled notation). The value of $\hat{\bb}$ is available in practice as it is the outcome of the regression. We typically only have access to the empirical covariance matrix from which to infer the covariate correlation matrix $\A$. A possible solution for this problem is to use the link between the empirical and population level eigenvalue distributions  in the Marchenko-Pastur equation \cite{marchenko1967distribution}.  The population spectrum can be estimated from its empirical counterpart in \cite{el2008spectrum}, by applying convex optimisation to the inverted Marchenko-Pastur equation. This method is an improvement on naively using the sample eigenvalue spectrum as an estimator of its population counterpart when $p > N$. 

\black
 
 There are many directions for extension of the present line of research. 
For instance,  time-to-event data, whether from observational studies of clinical trials, are typically censored. Censoring may reflect the impact of competing risks, patients withdrawing from studies, or finite study durations.  The incorporation of  censoring into our theory is an obvious next research target, together with investigation of regimes where the risk score are no longer Gaussian distributed. 
Finally, the overfitting measure in \eref{eq:replica1} is quite general, and can be applied to many other survival analysis models \cite{klein2006survival}. Equally, the theory developed in this paper is directly applicable to time-to-event studies outside medical data such as credit risk analysis.
\vsp

\noindent{\bf Acknowledgements}\\[2mm]
MS is supported by the Biotechnology and Biological Sciences Research Council (award 1668568) and GSK Ltd. He would like to thank the Isaac Newton Institute for Mathematical Sciences for support and hospitality during the programme `Statistical Scalability' when some of the work in this paper was undertaken. Both authors are grateful to Alexander Mozeika and Fabian Aguirre Lopez for constructive discussions.

\section*{References}
\bibliographystyle{unsrt}
%\bibliography{bib_replica}

\appendix 

\section{Self-averaging with respect to true associations}
\label{app:self_averaging}

Here we investigate properties of random variables of the form $\mathcal{R}=p^{-1}\bb^0\cdot {\bf P}\bb^0$ in the limit $p\to\infty$, where the   true association vectors $\bb^0=\{\beta_\mu^0\}$ are drawn randomly from some distribution $p(\bb^0)$ and ${\bf P}$ is a fixed symmetric positive definite $p\times p$ matrix, which is independent of $\bb^0$. In particular, we wish to determine under which conditions $\mathcal{R}$ will be self-averaging, i.e. $\lim_{p\to\infty}\bra \mathcal{R}\ket>0$ exists, and $\lim_{p\to\infty}[\bra \mathcal{R}^2\ket-\bra \mathcal{R}\ket^2]=0$. Brackets will in this Appendix denote averaging over $p(\bb^0)$, and we will write the eigenvalue distribution of ${\bf P}$ as $\varrho(\lambda)$. We make the following assumptions:\footnote{Assuming distinct variances for each $\beta_{\mu}^0$ complicates various equations but ultimately leads to similar final conditions on the eigenvalue spectrum of $\A$.}
\begin{enumerate}
\item The $\{\beta_\mu^0\}$ are independent and identically distributed, i.e. $p(\bb^0) = \prod_{\mu=1}^p p(\beta_{\mu}^0)$. 
\item $p(\bb_\mu^0)$ is symmetric in $\bb_\mu^0$,  with finite second and fourth order moments. 
  \end{enumerate}
 In view of our earlier definition $S^2=\lim_{p\to\infty}p^{-1}(\bb^0)^2\!$, we must identify $\bra  (\bb_\mu^0)^2\ket=S^2$. 
 We will write $\Sigma=\bra  (\bb_\mu^0)^4\ket$. 
 It then follows that
\begin{eqnarray}
\label{eq:expectation_R}
\hspace*{-20mm}
\lim_{p\to\infty}\langle \mathcal{R} \rangle
&=&
\lim_{p \rightarrow \infty} \frac{1}{p} \sum_{\mu\nu=1}^p   \bra\beta^0_{\mu}\beta^0_{\nu} \ket P_{\mu\nu} 
= S^2 \lim_{p\to\infty}\int\!\rmd\lambda~\varrho(\lambda)\lambda
\\
\hspace*{-20mm}
\lim_{p\to\infty}\langle \mathcal{R}^2 \rangle
&=& 
 \lim_{p \rightarrow \infty} \frac{1}{p^2} \sum_{\mu\nu\kappa\tau=1}^p \!
  \big\langle \beta^0_{\mu} \beta^0_{\nu} \beta^0_{\kappa} \beta^0_{\tau}  \big\rangle P_{\mu\nu} P_{\kappa \tau} 
  \nonumber \\
  \hspace*{-20mm}
  &=& 
\lim_{p \rightarrow \infty} \frac{1}{p^2} \Big\{
S^4{\rm Tr}^2({\bf P}) +2S^4{\rm Tr}({\bf P}^2)+(\Sigma\!-\!3S^4) 
\sum_{\mu=1}^p (P_{\mu\mu})^2 \Big\}
  \nonumber \\[-1mm]
  \hspace*{-20mm}
  &\leq & \Big(\lim_{p\to\infty}\langle \mathcal{R} \rangle\Big)^2+
\lim_{p \rightarrow \infty} \frac{1}{p} 
(\Sigma\!-\!S^4)\int\!\rmd\lambda~\varrho(\lambda)\lambda^2
\end{eqnarray}
We conclude that $\mathcal{R} $ will be self-averaging in the limit $p\to\infty$ if  $\lim_{p\to\infty}\int\!\rmd\lambda~\varrho(\lambda)\lambda$ exists and $\lim_{p \rightarrow \infty} p^{-1}
\int\!\rmd\lambda~\varrho(\lambda)\lambda^2=0$. Equivalently,
\begin{eqnarray}
\lim_{p\to\infty}\frac{1}{p}\sum_{\mu=1}^p P_{\mu\mu}\in \Re~~~~~~{\rm and}~~~~~~\lim_{p\to\infty}\frac{1}{p^2}\sum_{\mu\nu=1}^p P^2_{\mu\nu}=0
\end{eqnarray}
The two relevant quadratic expression for which we seek to demonstrate self-averaging are the following:
\begin{itemize}
\item
Application to ${\bf P}=\A$ tells us that if $\lim_{p\to\infty}\bra a\ket\in \Re$ and $\lim_{p\to\infty}p^{-1}\bra a^2\ket=0$ (i.e. the covariate correlations are not excessive), then
\begin{eqnarray}
\tilde{S}^2&=&\lim_{p\to \infty}\frac{1}{p}\bb^0\cdot\A\bb^0=S^2\bra a\ket
\end{eqnarray}
\item
Our second application is to the following matrix, in which the vectors $\{\vv^\mu\}$ are the orthogonal and normalised eigenvectors  of $\A$, with eigenvalues $a_\mu$:
\begin{eqnarray}
P_{\mu\nu}&=& \sum_{\rho=1}^p \frac{a_\rho^2 v_\mu^\rho v_\nu^\rho}{2\eta\gamma+ga_\rho}
\end{eqnarray}
Here we find, anticipating that $g>0$ and using $\eta\gamma>0$, 
\begin{eqnarray}
\hspace*{0mm}
\frac{1}{p}\sum_{\mu=1}^p P_{\mu\mu}&=&  \frac{1}{p}\sum_{\rho=1}^p \frac{a_\rho^2 }{2\eta\gamma\!+\!ga_\rho}\leq \frac{\bra a\ket}{g}
\\
\hspace*{0mm}
\frac{1}{p^2}\sum_{\mu\nu=1}^p P^2_{\mu\nu}&=& 
 \frac{1}{p^2}\sum_{\rho\rho^\prime\mu\nu=1}^p \frac{a_\rho^2 a_{\rho^\prime}^2 v_\mu^\rho v_\nu^\rho  v_\mu^{\rho^\prime} v_\nu^{\rho^\prime}}{(2\eta\gamma\!+\!ga_\rho)(2\eta\gamma\!+\!ga_{\rho^\prime})}
  \nonumber
  \\
 &=&
  \frac{1}{p^2}\sum_{\rho=1}^p \frac{a_\rho^4}{(2\eta\gamma\!+\!ga_\rho)^2}
 \leq \frac{\bra a^2\ket}{pg}
\end{eqnarray}
We conclude, provided $g>0$, that the same two conditions on $\A$ that guarantee self-averaging of $\tilde{S}^2$ for $p\to\infty$
will also imply self-averaging here:
\begin{eqnarray}
\lim_{p\to\infty} \frac{1}{p} \sum_{\rho=1}^p \frac{a_\rho^2 (\bb^0\!\cdot \vv^\rho)^2 }{2\eta\gamma+ga_\rho}&=& 
\Big\bra \frac{S^2a^2}{2\eta\gamma\!+\!ga}\Big\ket
\end{eqnarray}
\end{itemize}
Thus,  for our RS theory to be self-averaging with respect to the realisation of the true association vector $\bb^0$  (given our mild assumptions on the distribution from which $\bb^0$ is drawn), it is sufficient  that average and width of the eigenvalue distribution $\varrho(a)$ of the covariate correlation matrix $\A$ remain finite in the limit $p\to\infty$.

\end{document}